\documentclass[twocolappendix]{emulateapj}  

\usepackage{color}
\usepackage{epsfig}
\usepackage{subfigure,verbatim}
\usepackage[colorlinks,urlcolor=blue,citecolor=blue,linkcolor=blue]{hyperref}

\newcommand{\ud}{{\rm d}} 

\newcommand{\grad}{\nabla}
\newcommand{\cross}{\times}
\newcommand{\bmath}[1]{\mbox{\boldmath{$#1$}}}

\newcommand{\bi}{\begin{itemize}}
\newcommand{\ei}{\end{itemize}}

\newcommand{\ex}[1]{10^{-#1}}

\newcommand{\eq}[1]{Eq.~(\ref{eq:#1})}
\newcommand{\eqn}[1]{(\ref{eq:#1})}
\newcommand{\fig}[1]{Fig.~\ref{fig:#1}}
\newcommand{\sect}[1]{Section \ref{sec:#1}}
\newcommand{\app}[1]{Appendix \ref{sec:#1}}
\newcommand{\dg}{\Delta\gamma}
\newcommand{\be}{\begin{eqnarray}}
\newcommand{\ee}{\end{eqnarray}}
\newcommand{\rli}{\,r_{L,i}}
\newcommand{\qt}{(1+q\,t)}
\newcommand{\bavg}{\langle \bmath{B}\rangle}

\newcommand{\dbperp}{\delta B_\perp}
\newcommand{\perpe}{{e\perp}}
\newcommand{\pare}{{e\parallel}}
\newcommand{\perpi}{{i\perp}}
\newcommand{\pari}{{i\parallel}}

\def\L{\bmath{{L}}}
\def\lab{_{\rm L}}
\def\bvec{\bmath{B}}
\def\bhat{\bmath{\hat{b}}}
\def\evec{\bmath{E}}

\def\ra{\rangle}
\def\la{\langle}
\def\powi{{0.4}}
\def\powe{{0.5}}
\def\coeffi{{0.65}}
\def\coeffe{{0.55}}

\begin{document}
\title{Electron Heating by the Ion Cyclotron Instability  in Collisionless Accretion Flows.\\
I. Compression-Driven Instabilities and the Electron Heating Mechanism}
\author{Lorenzo Sironi$^{1,2}$ and Ramesh Narayan$^1$}
\affil{$^1$Harvard-Smithsonian Center for Astrophysics, 
60 Garden Street, Cambridge, MA 02138, USA
\\
$^2$NASA Einstein Postdoctoral Fellow}
 \email{E-mail: lsironi@cfa.harvard.edu; rnarayan@cfa.harvard.edu}
 
 \shorttitle{Electron Heating by the Ion Cyclotron Instability in Collisionless Accretion Flows}
\shortauthors{L. Sironi and R. Narayan}
\begin{abstract}
In systems accreting well below the Eddington rate, such as the central black hole in the Milky Way (Sgr A$^*$), the plasma in the innermost regions of the disk is believed to be collisionless and two-temperature, with the ions substantially hotter than the electrons. Yet, whether a collisionless faster-than-Coulomb energy transfer mechanism exists in two-temperature accretion flows is still an open question. We study the physics of electron heating during the growth of ion velocity-space instabilities, by means of multi-dimensional fully-kinetic particle-in-cell (PIC) simulations. A background large-scale compression --- embedded in a novel form of the PIC equations --- continuously amplifies the field. This constantly drives  a pressure anisotropy $P_\perp>P_\parallel$, due to the adiabatic invariance of the particle magnetic moments. We find that, for ion plasma beta values $\beta_{0i}\sim 5-30$  appropriate for the midplane of low-luminosity accretion flows (here, $\beta_{0i}$ is the ratio of ion thermal pressure to magnetic pressure), mirror modes dominate if the electron-to-proton temperature ratio is $T_{0e}/T_{0i}\gtrsim 0.2$, whereas for $T_{0e}/T_{0i}\lesssim 0.2$ the ion cyclotron instability triggers the growth of strong Alfv\'en-like waves, that pitch-angle scatter the ions to maintain marginal stability. We develop an analytical model of electron heating during the growth of the ion cyclotron instability, which we validate with PIC simulations. We find that for cold electrons ($\beta_{0e}\lesssim 2 \,m_e/m_i$, where $\beta_{0e}$ is the ratio of electron thermal pressure to magnetic pressure), the electron energy gain is controlled by the magnitude of the E-cross-B velocity induced by the ion cyclotron waves. This term is independent of the initial electron temperature, so it provides a solid energy floor even for electrons starting with extremely low temperatures. On the other hand, the electron energy gain for $\beta_{0e}\gtrsim 2 \,m_e/m_i$ --- governed by  the conservation of the particle magnetic moment in the growing fields of the instability --- is proportional to the initial electron temperature, and it scales with the magnetic energy of ion cyclotron waves. Our results have implications for two-temperature accretion flows as well as the solar wind and intracluster plasmas.
\end{abstract}


\section{Introduction}\label{sec:intro}
At mass accretion rates less than about 1\% of the Eddington rate,
accretion flows around black holes switch from the usual thin
accretion disk \citep{ss73,nt73,fkr02} to a so-called
advection-dominated accretion flow
\citep[or ADAF;][]{ny94,ny95a,ny95b,abr+95,nmq98,nm08,yuan_narayan_14}. 
ADAFs are relevant for understanding the
hard and quiescent state of black hole X-ray binaries
\citep{nmy96,nbm97,esin+97,esin+98}, as well as accretion in
low-luminosity active galactic nuclei
\citep{fab_rees,lasota+96,rey+96,dimat+00}, including the
ultra-low-luminosity source Sagittarius A$^*$ (Sgr A$^*$) at our
Galactic Center
\citep{nym95,nar+98,yuan+03,xu+06,moscibrodzka+12}. 

The key feature of an ADAF is that the accreting gas heats up close to
the virial temperature, causing the accretion flow to puff up into a
geometrically thick configuration, and the plasma to become
optically thin.  Because of the very low gas density, the timescale for electron and ion Coulomb collisions is much longer than the inflow time in the disk, i.e., the plasma is
largely collisionless. Furthermore, at distances less than a few hundred
Schwarzschild radii from the black hole ($R_S\equiv 2\,GM_\bullet/c^2$ is the Schwarzschild radius, where $M_\bullet$ is the black hole mass), the rate of Coulomb collisions is so low that ions and electrons are thermally decoupled and the plasma is two-temperature, with
the ions substantially hotter than the electrons
\citep{ny95b,yuan+03}. 


The two-temperature nature of the gas in ADAFs is a generic
prediction because (\textit{i}) electrons radiate much more efficiently than
ions, (\textit{ii}) coupling between ions and electrons via Coulomb collisions
is inefficient at the low densities expected in the innermost regions of ADAFs, and (\textit{iii}) compressive heating favors
non-relativistic ions over relativistic electrons. 
Despite these strong reasons, the plasma could still be driven to a
single-temperature state if there were additional modes of electron heating (beyond electron-ion Coulomb collisions). 

A number of processes have been suggested that might heat electrons in accretion flows, including direct heating by plasma waves or magnetohydrodynamic (MHD)
turbulence \citep{begelman88,quataert98,blackman99,medvedev00}
and  reconnection \citep{bisno97,quataertgruzinov99}.\footnote{Shocks are commonly invoked as a mechanism for particle heating and acceleration in the Universe \citep[see, e.g.,][for a study of electron acceleration in low Mach number shocks, with parameters relevant for our Galactic Center]{guo_14a,guo_14b}. Yet, fluid motions in accretion flows are generally subsonic, so shocks are not expected to occur.}  Recently, shearing
box simulations of accretion flows in the so-called kinetic MHD
framework \citep{sharma06,sharma07}, which incorporates the evolution of a pressure tensor in
the MHD equations of the plasma, have suggested an additional
mechanism for particle heating in ADAFs. An increase in the magnetic
field due to compression or shear, coupled with the adiabatic invariance of
particle magnetic moments, gives rise to an anisotropy  in the plasma temperature (with respect
to the local magnetic field). The resulting anisotropic pressure stress can contribute to angular momentum transport and particle heating in low-luminosity accretion flows \citep{quataert_02,sharma06,riquelme12,hoshino_13}.

In a collisional plasma, the magnitude of the pressure anisotropy is set by the frequency of Coulomb collisions. In contrast, in a collisionless plasma, the pressure anisotropy is regulated by the growth of small-scale instabilities that violate adiabatic invariance (most importantly, the mirror, ion cyclotron, firehose and electron whistler instabilities). These instabilities pitch-angle scatter the particles and tend to
isotropize the particle velocity distribution. Anisotropy-driven instabilities have been shown to play an important role in  the solar wind \citep{kasper_02,kasper_06,bale_09,maruca_11,maruca_12,matteini_07,matteini_13,cranmer_09,cranmer_12} and in the intracluster medium \citep{scheko_05,lyutikov_07,santos-lima_14}. The same
instabilities will regulate the heating of electrons and ions
in collisionless accretion flows. While the linear growth of these instabilities can be properly described with analytical tools  \citep[see][for a review of the kinetic stability of plasmas
with anisotropic pressure]{gary_book}, their saturation and non-linear evolution requires self-consistent particle-in-cell (PIC) simulations.

In this work, we investigate the physics of electron heating during the growth of anisotropy-driven ion instabilities by means of one- and two-dimensional (1D and 2D) fully-kinetic PIC simulations. In contrast to previous studies, we do not consider the initial value problem of the evolution of a prescribed pressure anisotropy \citep[see][for a review]{gary_book}. Instead, we self-consistently induce the growth of a temperature anisotropy by continuously amplifying the mean magnetic field in our computational domain. The increase in magnetic field, coupled to the adiabatic invariance of the particle magnetic moments, constantly drives a temperature anisotropy $T_\perp> T_\parallel$ with respect to the background magnetic field. This setup allows us to study the long-term non-linear evolution of the relevant instabilites, and not only their initial exponential stage. In our setup, the increase in magnetic field is driven by {\it compression}, mimicking the effect of large-scale compressive motions in ADAFs.  However, our results hold regardless of what drives the field amplification, so they can be equally applied to the case where velocity-space instabilities are induced by incompressible {\it shear} motions.

We have modified the standard PIC equations (i.e., Maxwell's equations and the Lorentz force) to account self-consistently for the overall compression of our system.\footnote{Our method can also be applied to an expanding plasma, but in this work we only study compressing systems.} This technique is complementary to the method described in  \citet{riquelme12} and employed in \citet{riquelme_14}, which is appropriate for incompressible shear flows (see also \citealt{kunz_14}, for hybrid simulations of shear-driven velocity-space instabilities). Hybrid simulations of compression-driven systems have been widely employed to study  {\it ion} velocity-space instabilities, most commonly for solar wind applications \citep[e.g.,][]{hellinger_05}. However, hybrid models --- that treat the ions as kinetic particles, but the electrons as a massless charge-neutralizing isotropic fluid --- cannot describe self-consistently the physics of {\it electron} velocity-space instabilities, nor the efficiency of electron heating during the growth of ion-scale instabilities. Fully-kinetic PIC simulations  in electron-positron plasmas (so, with mass ratio $m_i/m_e=1$) suffer from the same limitations \citep{riquelme_14}.\footnote{\citet{riquelme_14} also performed a few simulations with $m_i/m_e=4$, to show that the ion physics does not depend on the mass ratio. Yet, such values of $m_i/m_e$ are likely too small to properly capture the physics of electron-scale instabilities. \citet{riquelme_14} only explored the case where ions and electrons have equal temperatures, whereas in our work we study the dependence on the electron-to-proton temperature ratio.} A comprehensive investigation of the physics of electron heating induced by ion velocity-space instabilities is the main subject of this work, where we employ fully-kinetic PIC simulations with ion-to-electron mass ratios $m_i/m_e\gg1$.

We find that, for values of ion plasma beta $\beta_{0i}\sim 5-30$, similar to those expected in the midplane of low-luminosity accretion flows \citep[e.g.,][]{sadowski_13} (here, $\beta_{0i}$ is the ratio of ion thermal pressure to magnetic pressure), the nature of the dominant anisotropy-driven ion instability changes as a function of the electron-to-ion temperature ratio. If $T_{0e}/T_{0i}\gtrsim 0.2$, the mirror instability dominates, as revealed by its characteristic pattern of non-propagating modes whose wavevector is oblique to the background field \citep[e.g.,][]{hasegawa_69,southwood_93,kivelson_96}. In contrast, if $T_{0e}/T_{0i}\lesssim 0.2$, it is the ion cyclotron mode \citep[e.g.,][]{gary_76,gary_book,hellinger_06} that controls the evolution of the system during the course of compression. Since the wavevector of the ion cyclotron instability is aligned with the mean magnetic field, the relevant physics can be conveniently studied by means of 1D simulations, with the box aligned with the ordered field. Thanks to the greater number of computational particles per cell allowed by 1D simulations, as opposed to 2D, the efficiency of electron heating by the ion cyclotron instability can be reliably estimated. We develop an analytical model to describe the physics of electron heating during the growth of the ion cyclotron instability, and we successfully validate our model with 1D PIC simulations in the two extreme cases of cold ($\beta_{0e}\lesssim 2\, m_e/m_i$, where $\beta_{0e}$ is the electron beta, i.e., the ratio of electron thermal pressure to magnetic pressure) and warm ($\beta_{0e}\gtrsim 2\, m_e/m_i$) electrons. In a forthcoming paper (hereafter, Paper II), we will investigate how the electron heating efficiency by the ion cyclotron instability depends on  flow conditions.

This work is organized as follows. In \sect{setup} we describe the basic setup of our simulations, deferring to \app{method} a detailed derivation of the relevant equations of our PIC method in a compressing (or expanding) box. In \sect{struct}, we present 1D and 2D simulations of compression-driven instabilities, emphasizing the role of the electron-to-ion temperature ratio $T_{0e}/T_{0i}$ in determining the nature of the dominant mode. In \sect{heating}, we present our analytical model for electron heating during the growth of the ion cyclotron instability, and we validate our model for two representative cases. We summarize our results in \sect{summary} and we emphasize the astrophysical implications of our findings.

\begin{figure}[tbp]
\begin{center}
\includegraphics[width=0.5\textwidth]{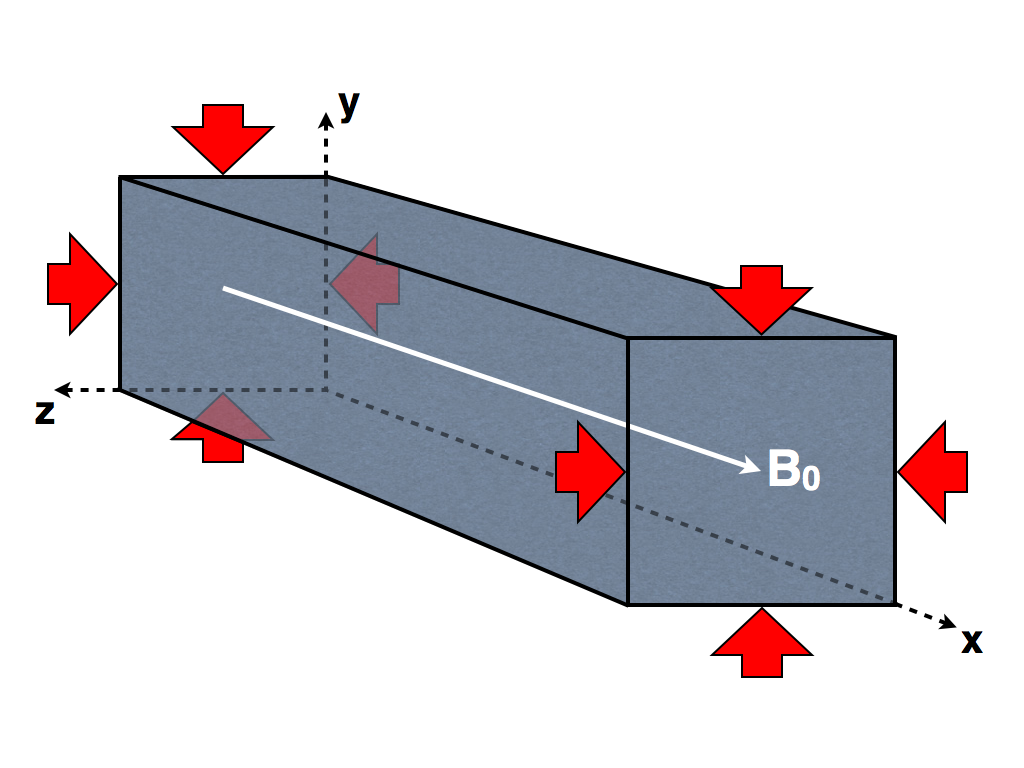}
\caption{Simulation setup. The magnetic field is initialized along the $x$ direction, and the box is compressed along the $y$ and $z$ directions, perpendicular to the magnetic field.}
\label{fig:setup}
\end{center}
\end{figure}

\section{Simulation Setup}\label{sec:setup}
We investigate compression-driven instabilities in collisionless accretion flows by means of fully-kinetic PIC simulations. We have modified the three-dimensional (3D) electromagnetic PIC code TRISTAN-MP \citep{buneman_93,spitkovsky_05,ssa_13,ss_14,sironi_giannios_14} to account for the effect of an overall compression of the system. So far, a compressing (or expanding) computational domain has been employed to study the behavior of astrophysical plasmas in magnetohydrodynamic simulations \citep{velli_93,velli_96,delzanna_12} or hybrid models with kinetic ions and fluid electrons \citep{hellinger_03b,hellinger_03,hellinger_05, liewer_01,matteini_12}. Our model, which we describe in detail in \app{method}, accounts for the kinetic physics of both ions and electrons, so it provides the first implementation of the fully-kinetic PIC method in a compressing or expanding box. Our model is complementary to the technique described in  \citet{riquelme12} and employed in \citet{riquelme_14}, which is appropriate for incompressible shear flows.

We present here the main properties of our approach, deferring a detailed discussion to \app{method}. We  solve Maxwell's equations and the Lorentz force in the fluid {\it comoving} frame, which is related to the {\it laboratory} frame by a Lorentz boost. In the comoving frame, we adopt two sets of spatial coordinates, with the same time coordinate. The {\it unprimed} coordinate system in the fluid comoving frame has a basis of unit vectors, so it is the appropriate coordinate set to measure all physical quantities. Yet,  we find it convenient to re-define the unit length of the spatial axes in the comoving frame such that a particle subject only to compression or expansion stays at fixed coordinates. This will be our {\it primed} coordinate system in the fluid comoving frame. 

The location of a particle in the laboratory frame (identified by the subscript ``L'') is related to its position in the primed coordinate system of the fluid comoving frame by $\bmath{x}_{\rm L}= \bmath{L}\, \bmath{x}'$, where compression or expansion are described by the diagonal matrix 
\be\label{eq:lmatrix}
\L=\frac{\partial \bmath{x}}{\partial \bmath{x}'}=
\left(
\begin{array}{ccc}
 a_x \,& 0 \,& 0 \\
 0\,& a_y \,& 0 \\
 0 \,& 0 \,& a_z \\
\end{array}\right)~~~.
\ee
The coefficients $a_x$, $a_y$ and $a_z$, which vary with time but are independent of spatial location, will be specificied below. From the formalism described in \app{method}, we derive Maxwell's equations in the compressing (or expanding) box. The two evolutionary equations solved by the PIC method are, in the limit $|\dot{\L}\, \bmath{x}'|/c\ll1$ of non-relativistic compression speeds,
\be
\nabla'\cross(\L \bmath{E})  &=&-\frac{1}{c}\frac{\partial}{\partial t'}(\ell\, \bmath{L}^{-1} \bmath{B})\label{eq:basic1}~~,\\
\nabla'\cross(\L \bmath{B}) &=&\frac{1}{c}\frac{\partial}{\partial t'}(\ell\, \bmath{L}^{-1} \bmath{E})+\frac{4\pi}{c} \ell\, \bmath{J}'~~,\label{eq:basic2}
\ee
where the temporal and spatial derivatives pertain to the primed coordinate system (the reader is reminded that the primed and unprimed systems share the same time coordinate, so $\partial/\partial t'=\partial/\partial t$, whereas the spatial derivatives differ: $\nabla'=\L \,\nabla$). We define  $\evec$ and $\bvec$ to be the physical electromagnetic fields measured in the unprimed coordinate system. Here, $\ell=a_x a_y a_z$ is the determinant of $\L$. The electric current density $\bmath{J}'$ is computed by summing the contributions of individual particles, as described in \app{method}.  Due to the factors of $\ell$ and $\L$ appearing in Eqs.~\eqn{basic1} and \eqn{basic2}, the  Courant--Friedrichs--Lewy condition for numerical stability is more stringent in the case of a compressing box, as compared to the standard PIC method in a non-compressing box. Following the discussion in \app{method}, we choose the numerical speed of light to be $0.15$ cells/timestep, so that the   Courant--Friedrichs--Lewy condition is fulfilled throughout the timespan of our simulations.

The equations describing the motion of a particle with charge $q$ and mass $m$ can be written, still in the limit $|\dot{\L}\, \bmath{x}'|/c\ll1$ of non-relativistic compression speeds, as
\be
\frac{\ud \bmath{p}}{\ud t'}&=&-\,\dot{\L} \L^{-1} \bmath{p}+q\left(\evec+\frac{\bmath{v}}{c}\cross \bvec\right)\label{eq:basic3}~~,\\
\frac{\ud \bmath{x'}}{\ud t'}&=&\bmath{v}'\label{eq:basic4}~~,
\ee
where $\dot{\L}=\ud \L/\ud t$. The  physical momentum $\bmath{p}$ and velocity $\bmath{v}$ of the particle are measured in the unprimed coordinate system. Yet, the particle velocity  $\bmath{v}'$ entering \eq{basic4} should be evaluated in the primed coordinate system, where $\bmath{v}'=\L^{-1} \bmath{v}$. We remark that Eqs.~\eqn{basic3} and \eqn{basic4} hold for electrons and ions of arbitrary Lorentz factor (i.e., both non-relativistic, trans-relativistic and ultra-relativistic). This is in contrast to standard hybrid codes, where the particles are constrained to be non-relativistic.

In this work, we employ the PIC method described in \app{method} to study compression-driven instabilities in a magnetized plasma. A uniform ordered magnetic field $\bmath{B}_0$ is initialized along the $x$ direction, as shown in \fig{setup}, and the system is compressed along the $y$ and $z$ directions. More specifically, we choose
\be
a_x=1~~,~~a_y=a_z=\qt^{-1}~~,
\ee
 where $q$ is the prescribed compression rate. We prefer this choice, as compared to the alternative $a_y=a_z=1-q\,t$ adopted in the hybrid code of \citet{hellinger_05}, because it allows to evolve the system for a few characteristic compression timescales (to times $t\geq q^{-1}$). As a check, we have performed limited tests with $a_y=a_z=1-q\,t$, and confirmed that the evolution of compression-driven instabilities is similar to that seen in our nominal case $a_y=a_z=\qt^{-1}$.

As a result of compression, the extent of the physical system in the $y$ and $z$ directions will shrink as $\qt^{-1}$, so the density increases as $n=n_0 \qt^2$, where $n_0$ is the density at the initial time. From flux freezing, the ordered magnetic field grows as $\bvec=\bvec_0\qt^2$. In addition, as we discuss in \app{method}, the Lorentz force in \eq{basic3} dictates that, in the presence of the ordered field $\bvec$, the component of particle momentum aligned with the field does not change during compression, so $p_{\parallel}=p_{0\parallel}$, whereas the perpendicular momentum increases as $p_\perp=p_{0\perp}\qt$. This is consistent with the conservation of the first ($\mu\propto p_{\perp}^2/| \bvec|$) and second ($J\propto p_{\parallel} |\bvec|/n$) adiabatic invariants. Clearly, the compression of the system has the effect to increase both the magnetic and the particle energy content.

The method described in \app{method} can be applied to 1D, 2D or 3D computational domains. To follow the evolution of the system for longer times, and with sufficient accuracy to properly characterize the mechanism and efficiency of electron heating, we primarily employ 1D and 2D simulations. Yet, all three components of electromagnetic fields and particle velocities are tracked. We use periodic boundary conditions in all directions, assuming that the system is locally homogeneous, i.e., that gradients in the density or in the ordered field $\bvec_0$ are on scales larger than the box size. Due to our periodic boundaries, wave and particle convection outside of the box are neglected. Our choice of periodic boundary conditions is clearly allowed by the fact that both Maxwell's equations and the Lorentz force in the primed coordinate system (Eqs.\eqn{basic1}-\eqn{basic4}) have no explicit dependence on the spatial coordinates.

In \sect{struct}, we present at first 2D simulations of compression-driven instabilities in the $xy$ plane, using $128$ particles per species per cell (but we have confirmed that the results are the same with $512$ particles per species per cell). However, we find that, if the initial electron temperature is less than $\sim 20\%$ of the ion temperature, the wavevector of the dominant instability is aligned with the ordered magnetic field. It follows that the evolution of the dominant mode can be conveniently captured by means of 1D simulations with the computational box oriented along $x$.
With the same computational resources, in 1D we can greatly increase the number of computational particles per cell, as compared to 2D. In our 1D simulations, we typically employ 32,768 computational particles per species per cell, and we have tested that our results are the same when using up to 131,072 particles per species per cell. Such large values are of critical importance to suppress the spurious heating induced by the coarse-grained description of PIC plasmas \citep[e.g.,][]{melzani_13}, and so to reliably estimate the efficiency of electron heating by compression-driven ion instabilities.

We now describe the physical and numerical parameters of our simulations. The flow is characterized by the ion plasma beta, which is defined as
\be
\beta_{0i}=\frac{8 \pi n_0 k_B T_{0i}}{B_0^2}~~~,
\ee
where $n_0$ is the initial ion number density (equal to the electron number density) and the initial ion temperature is defined as $k_B T_{0i}=m_i\la \gamma_i v_i^2\ra/3$, where the average is computed over the ion distribution function.\footnote{Our definition of temperature, as the momentum flux in a given direction, is relativistically correct. It will be employed both for ions (which are generally non-relativistic) and for electrons (which can be non-relativistic, trans-relativistic or ultra-relativistic).}  For non-relativistic ions, the Larmor radius in the initial field is 
\be
r_{L,i}=\sqrt{\frac{3\,\beta_{0i}}{2}}\, \frac{c}{\omega_{0\rm pi}}~~,
\ee
where $c/\omega_{0\rm pi}=\sqrt{c^2 m_i/4\pi n_0 e^2}$
is the ion skin depth at the initial time. In this work, we focus on the case $\beta_{0i}=10-20$ most relevant for the innermost regions of accretion flows. In Paper II, we will explore a wider range of $\beta_{0i}$, from 5 up to 80. The electron thermal properties are specified via the electron plasma beta 
\be
\beta_{0e}=\frac{T_{0e}}{T_{0i}}\, \beta_{0i}~~.
\ee
 We vary the ratio $\beta_{0e}/\beta_{0i}$ from unity (i.e., equal temperatures) down to $\ex{3}$.

Initially, electrons and ions populate isotropic Maxwellian distributions. As a result of compression, the parallel temperature $k_B T_{\parallel}=m \la\gamma v_{\parallel} ^2\ra$ stays constant, whereas the perpendicular temperature $k_B T_{\perp}=m \la\gamma v_{\perp} ^2\ra$ increases, driving the system towards the threshold for anisotropy-induced instabilities. We stress that our definition of   parallel and perpendicular velocities is such that $\la v_{\parallel} ^2\ra=\la v_{\perp} ^2\ra$ for an isotropic plasma, so $\la v^2\ra=\la v_{\parallel }^2+2\,v_\perp^2\ra$. 
 
The magnetization is quantified by the Alfv\'en speed
\be
v_{A0i}=\frac{B_0}{\sqrt{4 \pi m_i n_0}}~~~,
\ee
so that the initial ion temperature equals $k_B T_{0i}=m_i\beta_{0i} v_{A0i}^2/2$. For non-relativistic ions, the ion cyclotron frequency is related to the ion plasma frequency by $\omega_{0ci}=(v_{A0i}/c)\, \omega_{0\rm pi}$. In this work, we fix $v_{A0i}/c=0.05$, but in Paper II we show that our results are the same when varying the Alfv\'en velocity from 0.025 up to 0.1.
 
 The ion cyclotron frequency  will set the characteristic  unit of time. In particular, we will scale the compression rate $q$ to be a fraction of the ion cyclotron frequency $\omega_{0ci}$. In accretion flows, we expect the ratio $\omega_{0 ci}/q$ to be much larger than unity ($\sim 10^7$, if $q$ is comparable to the local orbital frequency). Due to computational constraints, we will use values of $\omega_{0 ci}/q$ that are much smaller, yet still satisfying $\omega_{0 ci}/q\gg1$. In this work, we choose either $\omega_{0 ci}/q=50$ or 100, and in Paper II we demonstrate that the results presented here can be readily scaled to values $\omega_{0 ci}/q\gg1$ (there, we will show results up to $\omega_{0 ci}/q=3200$). We evolve the system up to a few compression timescales.
 
Since we are interested in capturing the efficiency of {\it electron} heating due to {\it ion} instabilities, we need to properly resolve the kinetic physics of both ions and electrons. We choose to resolve the initial electron skin depth $c/\omega_{0\rm pe}$ with 5 cells, but we have tested that our results are the same when using down to $c/\omega_{0\rm pe}=2.5$ cells and up to  $c/\omega_{0\rm pe}=20$ cells. As we discuss in \app{lecs}, the characteristic wavelength of the whistler instability --- driven by the electron anisotropy $A_e=\beta_\perpe/\beta_\pare-1$ --- is $\lambda_e\simeq 2\pi A_e^{-1/2} c/\omega_{\rm pe}$. In all the cases in which our system becomes unstable to the whistler mode, we have $A_e\lesssim1$. So, by resolving the electron skin depth, we are confident that we are also capturing the relevant physics of the electron whistler instability.

On the other hand, our computational domain needs to be large enough to include at least a few wavelengths of ion-driven instabilities. Both the mirror and the ion cyclotron mode grow on scales comparable to the ion Larmor radius $r_{L,i}\sim\sqrt{\beta_{0i} m_i/m_e}\, c/\omega_{0\rm pe}$. For computational convenience, in this work we use a reduced value of the ion-to-electron mass ratio, either $m_i/m_e=16$ or 64.\footnote{As we further demonstrate in Paper II, such values of $m_i/m_e$ are sufficient to separate the electron and ion physics (there, we extend our results up to $m_i/m_e=1024$).} Yet, in our analytical model we retain the explicit  dependence on the mass ratio. For $m_i/m_e=16$ and $\beta_{0i}=10$, which will be our reference case in 2D, we employ a square computational domain with $L_x=L_y=1536\,{\rm cells}\sim20\,r_{L,i}$, whereas $L_x=1536\,{\rm cells}\sim20\,r_{L,i}$ in 1D simulations. When changing $m_i/m_e$ or $\beta_{0i}$, we ensure that our computational domain is scaled such that it contains at least $\sim 10 \,r_{L,i}$, so that the dominant wavelength of ion-driven instabilities is properly resolved.


\section{Time Evolution of The Dominant Instability}\label{sec:struct}
In this Section, we present 1D and 2D simulations of compression-driven instabilities. We first discuss 2D simulations of our reference case with $\beta_{0i}=10$, $v_{A0i}/c=0.05$, $\omega_{0ci}/q=100$ and $m_i/m_e=16$, and we emphasize how the nature of the dominant mode changes as a function of the ratio of electron to ion temperatures. Oblique mirror modes dominate for $T_{0e}\sim T_{0i}$, but if $T_{0e}/T_{0i}\lesssim 0.2$, the exponential phase of growth is controlled by the ion cyclotron instability, whose wavevector is aligned with the mean field. We then proceed to study the main properties of the compression-driven ion cyclotron instability by means of 1D simulations.


\begin{figure*}[tbp]
\begin{center}
\includegraphics[width=\textwidth]{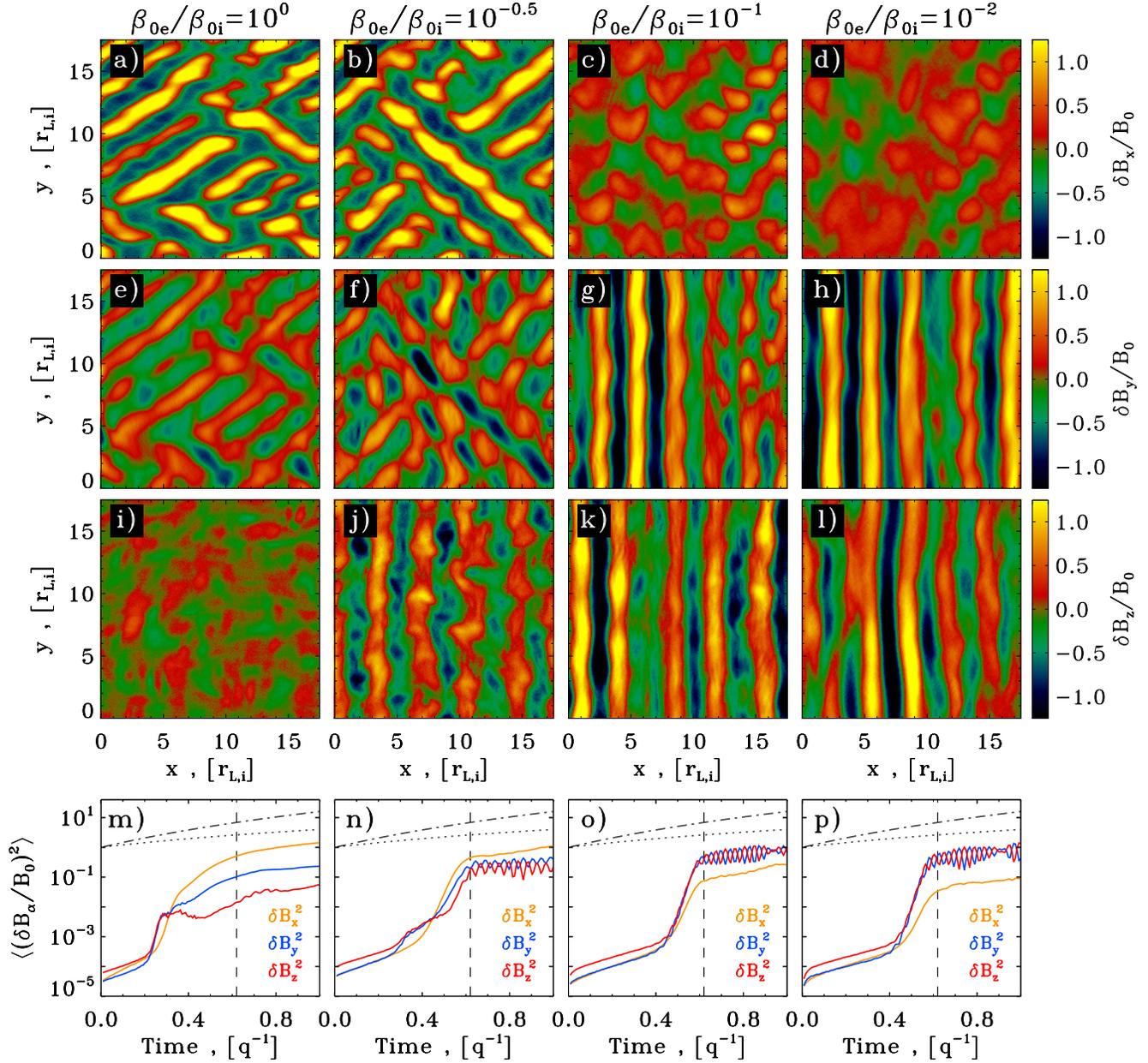}
\caption{2D spatial pattern and temporal evolution of compression-driven instabilities, for a representative case with $\beta_{0i}=10$, $v_{A0i}/c=0.05$, $\omega_{0ci}/q=100$ and $m_i/m_e=16$. Note that the nature of the dominant mode changes as a function of the ratio of electron-to-ion temperature: $\beta_{0e}/\beta_{0i}=1$ in the first column, $\ex{0.5}$ in the second, $\ex{1}$ in the third, and $\ex{2}$ in the last. The top three rows of panels show the 2D pattern of the turbulent magnetic fields, normalized to the initial field $B_0$ and projected along  $x$ (first row; this is the direction of the mean field),  $y$ (second row) and  $z$  (third row). The unit of length is the ion Larmor radius $r_{L,i}$. As the ratio $\beta_{0e}/\beta_{0i}=T_{0e}/T_{0i}$ decreases, longitudinal ion cyclotron waves in $\delta B_y$ and $\delta B_z$ dominate, at the expense of oblique mirror modes  in $\delta B_x$. The last row, with the compression timescale $q^{-1}$ as time unit, shows the temporal evolution of the average magnetic energy in the three directions, as indicated in the labels (orange for $\la\delta B_x^2\ra/B_0^2$, blue for $\la\delta B_y^2\ra/B_0^2$, red for $\la\delta B_z^2\ra/B_0^2$). For comparison, we also show $\qt^2$ (dotted black lines) and $\qt^4$ (dot-dashed black lines), where the latter describes the time evolution of the mean field energy.}
\label{fig:fluid2d}
\end{center}
\end{figure*}

\subsection{2D Results}\label{sec:struct2}
In \fig{fluid2d}, we examine our reference case with $\beta_{0i}=10$, $v_{A0i}/c=0.05$, $\omega_{0ci}/q=100$ and $m_i/m_e=16$, and we explore how the nature of the dominant mode depends on the ratio of electron to ion temperatures (or equivalently, on $\beta_{0e}/\beta_{0i}$). We vary the ratio of electron to proton temperature from $\beta_{0e}/\beta_{0i}=1$ down to $\ex{2}$, as indicated at the top of \fig{fluid2d}. We show the 2D pattern of magnetic field fluctuations in the $xy$ plane of the simulations. We present both the turbulent component along the mean field ($\delta B_\parallel= \delta B_x$,  in the top row) and the two components perpendicular to the field ($\delta B_y$ in the second row and $\delta B_z$ in the third row), normalized to the strength of the initial field $B_0$. The bottom row summarizes the temporal evolution of the magnetic energy density stored in the three components, averaged over the computational domain and normalized to $B_0^2/8\pi$ (orange for $\la \delta B_x^2\ra$, blue for $\la \delta B_y^2\ra$ and red for $\la \delta B_z^2\ra$).

In the case of equal temperatures of ions and electrons (leftmost column), we find that the 2D structure of the magnetic turbulence for $\beta_{0i}=\beta_{0e}=10$ is dominated by mirror modes. This is in agreement with earlier hybrid \citep[e.g.,][]{gary_76,gary_94,gary_97,gary_00,hellinger_05,kunz_14} and PIC studies \citep{riquelme_14}.\footnote{More precisely, earlier hybrid studies agreed that if the plasma beta is $\beta_{0i}=\beta_{0e}\gtrsim 7$, the dominant mode is the mirror instability, whereas the ion cyclotron instability dominates at smaller betas \citep[e.g.,][]{gary_94b}. However, the PIC study of shear-driven ion instabilities performed by \citet{riquelme_14} found that mirror modes dominate the non-linear evolution for all $\beta_{0i}=\beta_{0e}\gtrsim 1$. Our 2D simulations with $\beta_{0i}=\beta_{0e}=5$ are dominated by the mirror mode, in agreement with \citet{riquelme_14}.} As expected for the mirror instability, we find that the unstable fluctuations are non-propagating with their wavevector oblique to the background magnetic field \citep{hasegawa_69,southwood_93,kivelson_96}, as apparent from \fig{fluid2d}(a). We confirm that the mirror instability displays a spatial anti-correlation of magnetic field and density fluctuations, and that the turbulence in the parallel component is much stronger than in the perpendicular direction, i.e. $\la \delta B_x^2\ra\gg\la \delta B_y^2\ra, \,\la \delta B_z^2\ra$. This is apparent from the comparison of \fig{fluid2d}(a) with \fig{fluid2d}(e) and \fig{fluid2d}(i) (we employ the same color scale) and from the temporal evolution of the different turbulent components shown in \fig{fluid2d}(m). 

While our investigation of anisotropy-driven {\it ion} instabilities  in the case $\beta_{0i}=\beta_{0e}=10$ confirms the results of earlier studies, our fully-kinetic PIC method also allows us to capture the physics of {\it electron}-driven instabilities. We find that the bump at early times in $\la\delta B_y^2\ra$ and $\la\delta B_z^2\ra$ shown in \fig{fluid2d}(m) at $\simeq 0.25\,q^{-1}$ is generated by the electron whistler instability, which appears as counter-propagating transverse waves with $\la\delta B_y^2\ra \sim \la \delta B_z^2\ra\gg \la \delta B_x^2\ra$ having a wavelength comparable to the electron skin depth (we provide more details on the electron whistler instability in \app{lecs}). Hybrid studies, that treat the electrons as a fluid, or PIC simulations with $m_i/m_e=1$ (as in \citealt{riquelme_14}), cannot capture the growth of the electron whistler mode, which precedes the onset of the ion mirror instability.

The electron whistler instability also appears for $\beta_{0e}/\beta_{0i}=0.3$ in \fig{fluid2d}(n) at $q\,t\simeq0.35$, but with a smaller field amplitude relative to  the case $\beta_{0e}/\beta_{0i}=1$, due to the lower electron thermal content. Yet, the most dramatic difference shown by the case $\beta_{0e}/\beta_{0i}=0.3$ (in the second column), as compared to the case of equal temperatures (leftmost column), is the fact that the mirror mode is weaker (compare the orange lines between \fig{fluid2d}(m) and (n)), and a different instability pattern emerges in the perpendicular field components (see \fig{fluid2d}(f) and (j)). While the parallel fluctuations in \fig{fluid2d}(b) are still dominated by oblique modes, which are a signature of the mirror instability, the out-of-plane component $\delta B_z$ in \fig{fluid2d}(j) is controlled by a longitudinal mode, which coexists with the mirror fluctuations in the 2D plot of $\delta B_y$ in \fig{fluid2d}(f). This trend --- of weaker parallel fluctuations but stronger transverse waves  having a wavevector aligned with the mean field --- continues down to lower electron temperatures, as shown in the third column (with $\beta_{0e}/\beta_{0i}=10^{-1}$) and in the rightmost column (with $\beta_{0e}/\beta_{0i}=10^{-2}$). More precisely, the energy in the mirror-driven turbulence in $\la\delta B_x^2\ra$ is a factor of $\sim 3$ smaller than the energy in the perpendicular fluctuations at $\beta_{0e}/\beta_{0i}=10^{-1}$ (see \fig{fluid2d}(o)), and nearly an order of magnitude smaller than $\la\delta B_y^2\ra \sim\la \delta B_z^2\ra$ for $\beta_{0e}/\beta_{0i}=10^{-2}$ (see \fig{fluid2d}(p)).

The properties of the waves that dominate in the regime of cold electrons (i.e., $\beta_{0e}/\beta_{0i}\lesssim 0.2$)  point to the ion cyclotron mode, a kinetic instability that has maximum growth rate at nonzero frequencies below the ion cyclotron frequency, with propagation parallel to the background field \citep{kennel_66,davidson_75,schlickeiser_10}. As discussed by \citet{davidson_75}, the ion cyclotron mode survives in the limit of strongly magnetized electrons (or equivalently, $\beta_{0e}/\beta_{0i}\ll1$), provided that $\beta_{0i}\gtrsim1$, i.e., in the regime that we are investigating in this work. On the other hand, the mirror instability is non-resonant, so that the electron anisotropy contributes as much as the ion anisotropy to the growth of mirror fluctuations \citep[e.g.,][]{gary_92,gary_93,kuznetsov_12}. It follows that, if the electron thermal content is reduced, the electron contribution to the excitation of the mirror instability will also be suppressed, and the mirror modes will be weaker. Indeed, \citet{pokho_00} showed that the growth rate of the mirror instability, assuming that electrons and ions have the same temperature anisotropy (as expected in our case of driven compression), decreases significantly as the electron temperature is reduced from $T_{0e}=T_{0i}$ down to $T_{0e}\ll\,T_{0i}$. Similarly, \citet{remya_13} found that, for an undriven system with $T_{0e}=T_{0i}$, the inclusion of electron temperature anisotropy increases the proton anisotropy threshold of the ion cyclotron instability, which means that its development will be suppressed, whereas it decreases the anisotropy threshold for the mirror instability, facilitating its growth. We have tested the role of electron anisotropy by performing selected simulations in which the electrons are not affected by the compression of the system (i.e., we artificially set $\L$ to be the identity matrix in the electron equation of motion, so that the electrons remain isotropic). We find that, as compared to the case of compression-driven  anisotropic electrons, the mirror modes are suppressed even for $\beta_{0e}/\beta_{0i}\gtrsim 0.2$, in favor of ion cyclotron waves.\footnote{In the case where the electrons remain artificially isotropic, we do not observe the early growth of magnetic fluctuations that we have associated with the whistler instability, as expected.}

We have performed extensive convergence tests to check that the mirror instability is indeed suppressed --- in favor of the ion cyclotron instability --- as the electron-to-proton temperature ratio decreases below $T_{0e}/T_{0i}\lesssim 0.2$. We have confirmed our results when using four times as many particles per cell (1024 instead of 256, including both species), and we have also verified that our conclusions are unchanged for a box that is twice as large in each direction. Similarly, we have checked that the ion cyclotron mode still dominates when we choose a mass ratixo of $m_i/m_e=64$ (as compared to our standard choice $m_i/m_e=16$), an Alfv\'en velocity of $v_{A0i}/c=0.1$   (as compared to $v_{A0i}/c=0.05$) and a compression rate of $q/\omega_{0ci}=1/800$  (as compared to $q/\omega_{0ci}=1/100$). The dominance of the ion cyclotron mode over the mirror mode for $T_{0e}/T_{0i}\lesssim 0.2$ holds in two as well as in three dimensions. We have checked this by performing a 3D simulation with $\beta_{0i}=10$, $T_{0e}/T_{0i}=\ex{2}$, $v_{A0i}/c=0.05$,  $q/\omega_{0ci}=1/50$ and $m_i/m_e=16$ in a cubic box with 384 cells on each side, a spatial resolution of $2.5$ cells per skin depth and 32 computational particles per cell. We  found excellent agreement with its 2D counterpart.
In addition, we have confirmed that the critical threshold in $T_{0e}/T_{0i}$ (or equivalently, in $\beta_{0e}/\beta_{0i}$), that separates the mirror-dominated regime from the ion cyclotron-dominated regime, is only weakly dependent on the ion $\beta_{0i}$, ranging from $[T_{0e}/T_{0i}]_{\rm crit}\simeq0.3$ for $\beta_{0i}=5$ down to  $[T_{0e}/T_{0i}]_{\rm crit}\simeq0.1$ for $\beta_{0i}=80$. Finally, we have explicitly checked that our conclusions still hold when we adopt a different temporal evolution for the compression matrix $\L$ in \eq{lmatrix}, such that $\ddot{\L}=0$ ($a_y=a_z=1-q\,t$, as opposed to our standard choice $a_y=a_z=\qt^{-1}$).

As shown in \fig{fluid2d}, our conclusions about the relative role of mirror and ion cyclotron instabilities, as a function of the electron-to-ion temperature ratio, are valid during the exponential phase of growth and up to the characteristic compression time $\sim q^{-1}$. By evolving the system to longer times (up to $\sim 3\,q^{-1}$), we notice that the magnetic energy in parallel fluctuations grows faster than in the perpendicular components (compare the temporal evolution of the orange curve in \fig{fluid2d}(m) with the blue and red lines in \fig{fluid2d}(o) and (p)), so mirror-driven modes might dominate at late times, even for $T_{0e}/T_{0i}\lesssim 0.2$ (for the secular growth of mirror fluctuations, see \citealt{scheko_08,kunz_14,rincon_14}). However, the compression of the system might last for only $\sim q^{-1}$, in which case ion cyclotron waves will prevail. Also, as we argue in \sect{heating}, most of the electron heating happens during the phase of exponential growth, which for  $T_{0e}/T_{0i}\lesssim 0.2$ is clearly controlled by the ion cyclotron instability.


\begin{figure}[tbp]
\begin{center}
\includegraphics[width=0.45\textwidth]{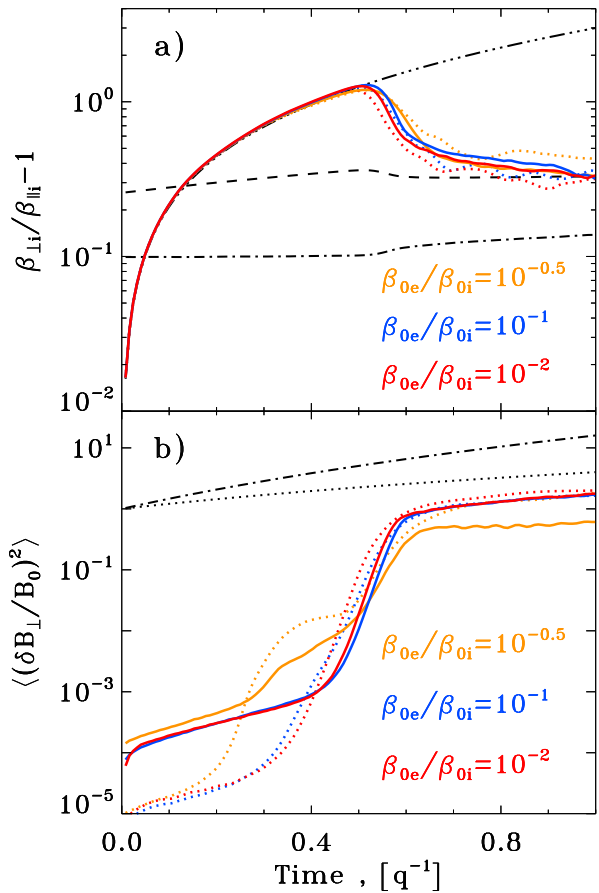}
\caption{Comparison of 2D and 1D results for a representative case with $\beta_{0i}=10$, $v_{A0i}/c=0.05$, $\omega_{0ci}/q=100$ and $m_i/m_e=16$, and different choices of the electron-to-ion temperature ratio (orange for $\beta_{0e}/\beta_{0i}=\ex{0.5}$, blue for $\ex{1}$, red for $\ex{2}$). 2D results are shown with the solid lines, 1D with the dotted lines, as a function of time in units of $q^{-1}$ (where $q$ is the compression rate). Top panel: ion anisotropy $A_i=\beta_\perpi/\beta_\pari-1$, together with the  time evolution $\qt^2-1$ expected in the case of compression alone (triple-dot-dashed black line). For comparison, we also show the threshold for the ion cyclotron (dashed black line)  and mirror (dot-dashed black line) instabilities, as in \eq{margi} and \eq{mirrori}, respectively (in the two equations, we calculate the right hand side using $\beta_{\perpi}$ and $\beta_\pari$ from the 1D run with $\beta_{0e}/\beta_{0i}=\ex{2}$). Bottom panel: magnetic energy in the perpendicular fields, in units of $B_0^2/8\pi$. The fact that at early times ($q\,t\lesssim 0.2$) the value of $\la\delta B_\perp^2\ra$ is lower in 1D than in 2D is because in 1D we can use a much greater number of particles per cell (typically, 65,536 in 1D as opposed to 256 in 2D), which reduces numerical noise. For comparison, we also show $\qt^2$ (dotted black line) and $\qt^4$ (dot-dashed black lines), where the latter describes the time evolution of the  mean field energy. For relatively cold electrons ($T_{0e}/T_{0i}\lesssim 0.2$), 1D simulations are perfectly adequate to capture the relevant physics, which is dominated by the ion cyclotron mode.
}
\label{fig:1d2d}
\end{center}
\end{figure}

\subsection{Comparison of 2D and 1D Results}
As we have argued above, if the initial electron temperature satisfies  $T_{0e}/T_{0i}\lesssim 0.2$, the exponential growth of compression-driven instabilities is controlled by the ion cyclotron mode, whose wavevector is aligned with the mean magnetic field. It follows that 1D simulations with the box oriented along the mean field (i.e., the $x$ direction, in our setup), should provide a good description of the overall physics. We test this hypothesis in \fig{1d2d}, where we present the temporal evolution of the ion anisotropy $A_i=\beta_\perpi/\beta_\pari-1$  (in the top panel) and of the magnetic energy in perpendicular fluctuations $\la\delta B_\perp^2\ra=\la \delta B_y^2\ra+ \la \delta B_z^2\ra$ (in the bottom panel), for 1D and 2D simulations (dotted and solid lines, respectively). The fact that at early times ($q\,t\lesssim 0.2$) the value of $\la\delta B_\perp^2\ra$ in \fig{1d2d}(b) is lower in 1D than in 2D is because, with the same computational resources, in 1D we can use a much greater number of particles per cell (typically, 65,536 as opposed to 256), which dramatically reduces numerical noise. In turn, this is essential to capture with sufficient accuracy the efficiency of electron heating. We explore temperature ratios $T_{0e}/T_{0i}=\beta_{0e}/\beta_{0i}$ of 0.3 (orange), $\ex{1}$ (blue) and $\ex{2}$ (red), for the same case as in \fig{fluid2d}, i.e., with fixed values of $\beta_{0i}=10$, $v_{A0i}/c=0.05$, $\omega_{0ci}/q=100$ and $m_i/m_e=16$.

For $T_{0e}/T_{0i}\lesssim 0.2$, our results are independent of the temperature ratio (blue lines for $T_{0e}/T_{0i}=\ex{1}$ and red lines for $T_{0e}/T_{0i}=\ex{2}$) and of the dimensionality of the computational domain (dotted for 1D and solid for 2D). This conclusion holds for the exponential rate of growth of magnetic energy (as measured from \fig{1d2d}(b)), for its saturation value, and for the temporal evolution in the secular phase at $q\,t\gtrsim 0.6$.

The temporal evolution of the ion anisotropy $A_i=\beta_\perpi/\beta_\pari-1$ in the top panel shows little difference between 1D and 2D, and little variation with temperature in the regime $T_{0e}/T_{0i}\lesssim 0.2$. The anisotropy initially increases, as a result of compression. Given the evolution of the parallel and perpendicular particle momenta discussed in \sect{setup} and in \app{method}, the anisotropy is expected to increase as $\qt^2-1$, in agreement with the triple-dot-dashed black line in \fig{1d2d}(a). At $q\, t\simeq 0.55 $, as a result of the growing turbulence induced by the ion cyclotron instability (shown in \fig{1d2d}(b)), the ion anisotropy is reduced by pitch-angle scattering \citep{kennel_66,mckean_92,mckean_94}. At late times, the anisotropy approaches the threshold of marginal stability for the ion cyclotron mode \citep[e.g.,][]{gary_76,gary_94,gary_94d,gary_97,gary_00,hellinger_06,yoon_12b,yoon_13}, namely,
\be\label{eq:margi}
\left[\frac{\beta_\perpi}{\beta_{\pari}}-1\right]_{\rm MS}=\frac{S_i}{\beta_{\pari}^{\alpha_i}}~~~.
\ee
Adopting $\alpha_i\simeq\powi$ (a common choice in the literature\footnote{However, \citet{isenberg_12,isenberg_13} argue that the threshold condition might be very different, if the particle distribution is not bi-Maxwellian.}), we find that our best-fit coefficient is $S_i\simeq\coeffi$. The condition of marginal stability in \eq{margi} is plotted in \fig{1d2d}(a) with a dashed black line, showing that it properly describes the late-time evolution of the ion anisotropy for $T_{0e}/T_{0i}\lesssim 0.2$, in both 1D and 2D.\footnote{From the temporal scalings of particle momenta discussed in \sect{setup},  we find that for non-relativistic ions $\beta_\perpi\propto {\rm const}$ and $\beta_{\pari}\propto\qt^{-2}$ before the instability grows.} In contrast, the ion anisotropy at late times is not consistent with the threshold of marginal stability of the mirror mode, which for cold electrons is \citep{stix_62,hasegawa_69,hellinger_07}
\be\label{eq:mirrori}
\left[\frac{\beta_{i\perp}}{\beta_{i\parallel}}-1\right]_{\rm MS}=\frac{1}{\beta_{i\perp}}~~~,
\ee
shown in \fig{1d2d}(a) with a dot-dashed black line.\footnote{As pointed out by \citet{riquelme_14}, the mirror threshold tends to be marginally higher than in \eq{mirrori}, for moderate values of $\omega_{0ci}/q$ (as opposed to the asymptotic limit $\omega_{0ci}/q\gg1$ appropriate for \eq{mirrori}).}

In summary, the ion physics for $T_{0e}/T_{0i}\lesssim 0.2$ is insensitive to the electron temperature and to the dimensionality of the simulation box (blue and red lines in \fig{1d2d}). Even for $T_{0e}/T_{0i}= 0.3$ (orange lines in \fig{1d2d}), we find that 1D simulations satisfactorily describe the physics of both ion-driven instabilities and electron-driven instabilities, despite artificially neglecting the oblique mirror modes. As regards to ion-driven instabilities, we find that the magnetic energy in perpendicular fluctuations saturates in 2D at a level that is similar as in 1D, differing by only a factor of two (compare solid and dotted orange lines in \fig{1d2d}(b) at $q \,t\gtrsim 0.6$). The agreement is even closer for the ion anisotropy (compare dotted and solid orange lines in \fig{1d2d}(a)), with 2D results showing a lower degree of anisotropy, possibly due to particle scattering by the combined effect of ion cyclotron and mirror modes (the latter being artificially excluded in 1D). 
As regards to the electron physics, we find that both 1D and 2D simulations for $T_{0e}/T_{0i}= 0.3$ show evidence of the electron whistler instability, which is responsible for the early growth of perpendicular magnetic energy at $q\,t\simeq0.3$ (dotted and solid orange lines in \fig{1d2d}(b)).

We conclude that, for relatively cold electrons ($T_{0e}/T_{0i}\lesssim 0.2$, and perhaps even for $T_{0e}/T_{0i}=0.3$), 1D simulations are perfectly adequate to capture the relevant physics, which is dominated by the ion cyclotron mode. In the following, we present 1D simulations of the compression-driven ion cyclotron instability, focusing on the mechanism and efficiency of electron heating.



\begin{figure*}[tbp]
\begin{center}
\includegraphics[width=1.05\textwidth]{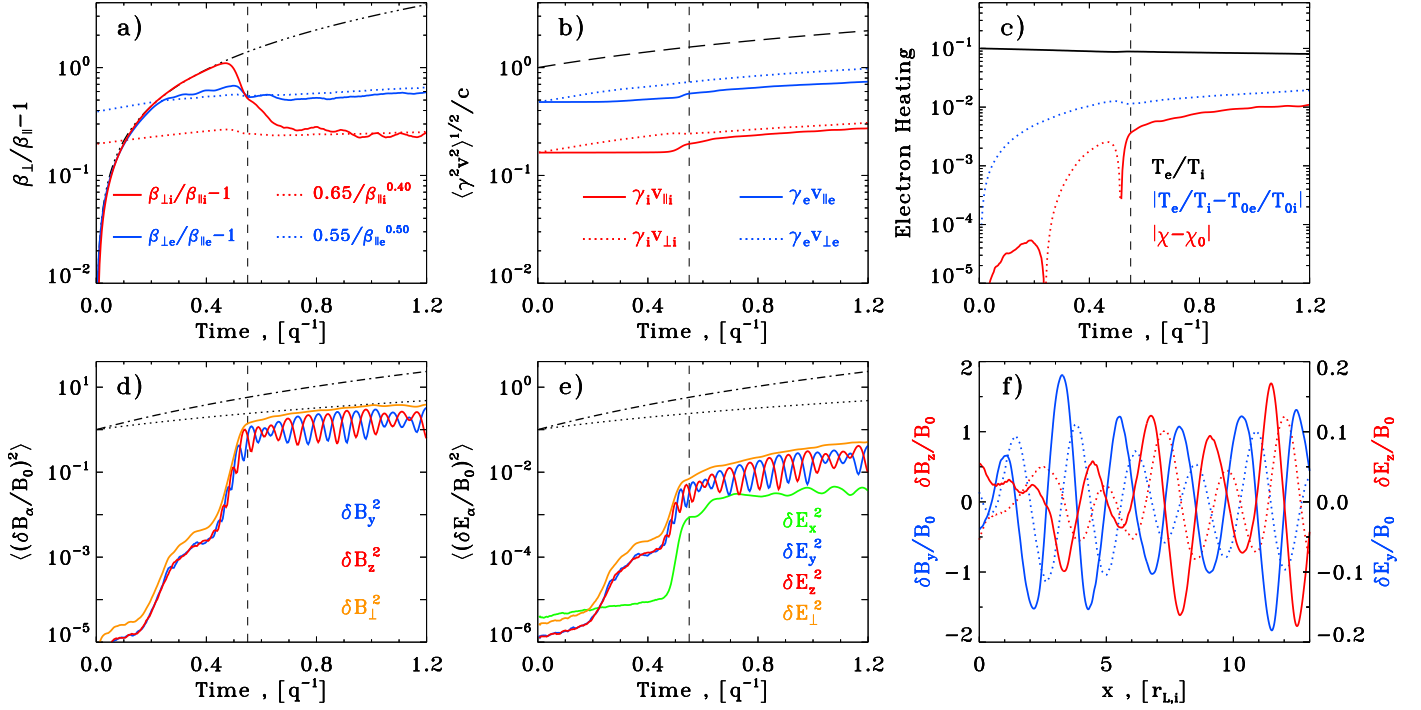}
\caption{Temporal and spatial development of compression-driven instabilities, from a representative 1D simulation with $\beta_{0i}=20$, $\beta_{0e}/\beta_{0i}=\ex{1}$, $v_{A0i}/c=0.05$, $\omega_{0ci}/q=100$ and $m_i/m_e=64$. We show the temporal evolution of the following quantities, with $q^{-1}$ as our time unit ($q$ being the compression rate): 
(a) Ion (red) and electron (blue) anisotropy $\beta_\perp/\beta_\parallel-1$, together with the threshold at marginal stability for the ion cyclotron instability (dotted red line, see \eq{margi}) and for the electron whistler instability (dotted blue line, see \eq{marge}). The triple-dot-dashed black line follows the track expected from compression alone.
(b) The mean momentum dispersion of ions (red) and electrons (blue), along (solid) or perpendicular (dotted) to the mean field. The dashed black line shows the expected evolution of the transverse component, due to compression alone (i.e., $\propto \qt$).
(c) Efficiency of electron heating or cooling. The black line is the temperature ratio $T_e/T_i$, the blue line its difference with respect to the initial value (solid if positive, dotted if negative), and the red line shows the $\chi$ parameter defined in \eq{chi}  relative to its initial value (solid if positive, dotted if negative).
(d) Energy density in magnetic fields, in units of $B_0^2/8\pi$ ($\la \delta B_y^2\ra$ in blue, $\la \delta B_z^2\ra$ in red, $\la \delta B_\perp^2\ra=\la \delta B_y^2+\delta B_z^2\ra$ in orange). For reference, we also plot $\qt^2$ (dotted black line) and $\qt^4$ (dot-dashed black line, as expected for the evolution of the mean field energy). 
(e) Energy density in electric fields, in units of $B_0^2/8\pi$ ($\la \delta E_x^2\ra$ in green, $\la \delta E_y^2\ra$ in blue, $\la \delta E_z^2\ra$ in red, $\la \delta E_\perp^2\ra=\la \delta E_y^2+\delta E_z^2\ra$ in orange). The dotted and the dot-dashed black lines are the same as in panel (d), apart from the normalization.
Finally, panel (f) shows --- at time $q\, t\simeq0.55$, as indicated with a vertical dashed line in all the other panels --- the spatial profile of the magnetic fields ($\delta B_y/B_0$ in solid blue and $\delta B_z/B_0$ in solid red, axis on the left) and of the electric fields ($\delta E_y/B_0$ in dotted blue and $\delta E_z/B_0$ in dotted red, axis on the right), as a function of the longitudinal coordinate $x$, measured in units of the ion Larmor radius $r_{L,i}$. 
}\label{fig:fluid1d}
\end{center}
\end{figure*}

\subsection{1d Results}\label{sec:struct1}
In this section, we describe the main properties of the ion cyclotron instability. This will  lay the foundations for the model of electron heating presented in \sect{heating}.

In the limit of weakly magnetized plasmas (namely, $v_{Ai}/c=\omega_{ci}/\omega_{\rm pi}\ll1$), the dispersion relation for ion cyclotron waves reads \citep[e.g.,][]{gary_book}
\be\label{eq:dispi}
\omega=\pm \,\omega_{ci}\,(1-\omega^2/k^2 v_{Ai}^2)~~~,
\ee
where the two signs indicate that the waves can move either along or opposite to the direction of the ordered field. For the compression-driven setup that we are studying here, the evolution at late times is such that the system remains roughly at the threshold of the ion cyclotron mode, as we  demonstrate below (but see also \fig{1d2d}(b)). The dominant wavevector at marginal stability is \citep{kennel_66,davidson_75,yoon_92,yoon_10}
\be\label{eq:ki}
k_i=\frac{\omega_{\rm pi}}{c}\frac{A_{i,\rm MS}}{\sqrt{A_{i,\rm MS}+1}}~~~,
\ee
where $A_{i,\rm MS}=[\beta_\perpi/\beta_\pari-1]_{\rm MS}$ is the ion anisotropy at the threshold of marginal stability for the ion cyclotron mode (\eq{margi}). For the range of $\beta_{i}\sim5-30$ appropriate for accretion flows, the anisotropy at marginal stability is $A_{i,\rm MS}\simeq\coeffi/\beta_{\pari}^{\powi}\ll1$. In this limit of weak anisotropy, the wavevector reduces to $k_i\simeq A_{i,\rm MS} \,\omega_{\rm pi}/c$. From \eq{dispi}, the frequency at marginal stability is\footnote{Hereafter, we indicate with $\omega_i$ the oscillation frequency of the ion cyclotron mode at marginal stability, whereas the ion cyclotron frequency is $\omega_{ci}$.}
\be\label{eq:omegai}
\omega_i=\pm\omega_{ci}\frac{A_{i,\rm MS}}{A_{i,\rm MS}+1}\simeq \pm A_{i,\rm MS} \,\omega_{ci}~~~,
\ee
so the phase speed is $\omega_i/k_i=v_{Ai}/\sqrt{A_{i,\rm MS}+1}\simeq v_{Ai}$.

In \fig{fluid1d}, we present the temporal evolution of compression-driven instabilities from a 1D simulation with $\beta_{0i}=20$, $\beta_{0e}/\beta_{0i}=\ex{1}$, $v_{A0i}/c=0.05$, $\omega_{0ci}/q=100$ and $m_i/m_e=64$. As we have discussed in \sect{setup}, in 1D simulations we can employ a number of computational particles per cell that is large enough to reliably capture the efficiency of electron heating.

As a result of compression, the anisotropy of both ions (solid red line in  \fig{fluid1d}(a)) and electrons (solid blue line in \fig{fluid1d}(a)) increases, due to the increase in the particle perpendicular momentum, whereas the parallel momentum is unchanged. This is shown in \fig{fluid1d}(b), where we plot the average momentum dispersion in the perpendicular (dotted) and parallel (solid) directions, for ions (red) and electrons (blue). The evolution of the anisotropy at early times  ($q\,t\lesssim 0.2$) follows the scaling $\beta_\perp/\beta_\parallel-1=\qt^2-1$ expected from compression alone (see \fig{fluid1d}(a), compare the solid lines with the triple-dot-dashed black line).

At $q\,t\simeq 0.2$, the electron anisotropy $A_e=\beta_\perpe/\beta_\pare-1$ exceeds the threshold of marginal stability for the electron whistler mode, which can be derived from linear theory \citep{gary_96,gary_06}
\be\label{eq:marge}
\left[\frac{\beta_\perpe}{\beta_{\pare}}-1\right]_{\rm MS}=\frac{S_e}{\beta_{\pare}^{\alpha_e}}~~~,
\ee
with $\alpha_e=\powe$ and our best-fit coefficient is $S_e\simeq\coeffe$. The condition of marginal stability for the electron whistler instability is indicated in \fig{fluid1d}(a) with a dotted blue line, and that for the ion cyclotron instability (\eq{margi}) with a dotted red line.

As  the electron anisotropy exceeds the threshold in \eq{marge}, we observe the growth of the electron whistler instability, whose properties are summarized in \app{lecs}. This leads to an exponential increase in the perpendicular magnetic energy (see the orange line in \fig{fluid1d}(d) at $q\,t \simeq 0.25$) and in the perpendicular electric energy (see the orange line in \fig{fluid1d}(e) at $q\,t \simeq 0.25$), until pitch-angle scattering by the magnetic fluctuations is so strong  that the electron anisotropy stays confined at marginal stability. Indeed, following the growth of the electron whistler instability, the electron anisotropy (solid blue line in \fig{fluid1d}(a)) never departs from the threshold of marginal stability (dotted blue line). 

The ions do not respond to (or participate in) the growth of the electron whistler instability. In fact, their anisotropy still follows the track $A_i=\qt^2-1$ expected from compression alone, until $q\,t\simeq 0.5$. At that point, the ion anisotropy exceeds the marginal condition in \eq{margi} by such a large amount that the system becomes unstable to the ion cyclotron mode.

We remark that the onset time of the ion cyclotron instability (and so, the maximum degree of ion anisotropy) is sensitive to the ratio $\omega_{0ci}/q$. As demonstrated in Paper II, as $\omega_{0ci}/q$ increases, the ion cyclotron instability appears earlier, and in the limit  $\omega_{0ci}/q\gg1$ it will grow soon after the ion anisotropy exceeds the threshold in \eq{margi}. Since this happens earlier than the onset of the electron whistler instability --- due to the fact that $S_e/\beta_{\pare}^{\alpha_e}>S_i/\beta_{\pari}^{\alpha_i}$, for the parameters chosen in \fig{fluid1d} --- the system might never exceed the threshold for the electron whistler instability, if the electron anisotropy is controlled by the ion cyclotron turbulence. The condition $S_e/\beta_{\pare}^{\alpha_e}\lesssim S_i/\beta_{\pari}^{\alpha_i}$ required for the whistler instability to precede the ion cyclotron growth can be recast as a lower limit on the electron-to-ion temperature ratio, 
\be
\frac{T_{0e}}{T_{0i}}\gtrsim \frac{0.7}{\beta_{0i}^{0.2}}\sim0.3-0.5~~~,
\ee
for the range of $\beta_{0i}\sim 5-30$ expected in accretion flows. This is already in the regime where oblique mirror modes cannot be neglected in the evolution of the system (see \sect{struct2}), so one needs to perform 2D simulations. For this reason, in the following we primarily focus on the ion cyclotron mode, deferring to Paper II a study of the impact of the electron whistler instability.

As a result of the ion cyclotron instability, the magnetic energy in the perpendicular component increases exponentially until $q\,t\simeq 0.55$ (which is indicated with a vertical dashed black line in \fig{fluid1d}), when the field reaches $\la \delta B_{\perp}^2\ra\sim B_0^2$ (orange line in \fig{fluid1d}(d)). In units of the mean field energy, which increases as $|\bavg|^2=|\bmath{B}_0|^2\qt^4$ as a result of compression (dot-dashed black line in \fig{fluid1d}(d)), we find that the magnetic energy in ion cyclotron waves at the end of the exponential phase reaches $\la \delta B_{\perp}^2\ra/|\bavg|^2\simeq 0.3$, for the value of $\beta_{0i}=20$ adopted in \fig{fluid1d}. A comparison of the orange line in \fig{fluid1d}(d) with the dotted blue line in \fig{1d2d}(b) --- having a different plasma beta, $\beta_{0i}=10$ --- shows that the saturation value of the ion cyclotron waves increases with the plasma beta.

In parallel with the growth of the magnetic field, the transverse electric energy (orange line in \fig{fluid1d}(e)) increases exponentially, saturating at $q\,t\simeq 0.55$ at a value of $\la \delta E_{\perp}^2\ra/|\bavg|^2\simeq \ex{3}$, or equivalently $\la \delta E_{\perp}^2\ra/\la \delta B_{\perp}^2\ra\simeq 3\times \ex{3}$. From Maxwell's equations, we expect that $\la \delta E_{\perp}^2\ra/\la \delta B_{\perp}^2\ra\sim (\omega_i/k_ic)^2\sim(v_{Ai}/c)^2$. Since $v_{A0i}/c=0.05$ in \fig{fluid1d}, we expect $\la \delta E_{\perp}^2\ra/\la \delta B_{\perp}^2\ra\simeq 2.5\times \ex{3}$, in excellent agreement with the results in \fig{fluid1d}.\footnote{The parallel electric energy $\la \delta E_x^2\ra$ (green line in \fig{fluid1d}(e)), which arises from charge fluctuations \citep{omidi_10}, is much smaller than $\la \delta E_\perp^2\ra$ in all the cases we have explored, and it will be neglected hereafter. In addition, parallel magnetic fluctuations are absent, since Maxwell's equations constrain $\la \delta B_x^2\ra$ to be zero in 1D simulations.}
 
Pitch-angle scattering by the growing ion cyclotron waves begins to reduce the ion anisotropy at $q\,t\simeq0.5$ (solid red line in \fig{fluid1d}(a), but see also the perpendicular and parallel ion momenta in \fig{fluid1d}(b)), when the energy in ion cyclotron waves is comparable to the ordered field, $\la \delta B_{\perp}^2\ra/|\bavg|^2\simeq 0.3$ (see \citealt{riquelme_14}, for similar conclusions in a system dominated by the mirror instability). Pitch-angle scattering brings the ion distribution back to the condition of marginal stability (indicated with a dotted red line in \fig{fluid1d}(a)).
At $q\,t\gtrsim 0.7$, while the ion anisotropy follows the threshold of marginal stability in \eq{margi}, the magnetic energy in perpendicular fluctuations grows secularly, with a temporal scaling that resembles $\la \delta B_{\perp}^2\ra\propto \qt^2$ (indicated with a dotted black line in \fig{fluid1d}(d)). At the same time, the electric energy in the transverse component grows faster, with $\la \delta E_{\perp}^2\ra\propto \qt^4$ (compare the orange line  with the dot-dashed black line in \fig{fluid1d}(e)). A temporal scaling with $\la \delta E_{\perp}^2\ra/\la \delta B_{\perp}^2\ra\propto \qt^2$ at late times is compatible with the fact that, from Maxwell's equations, $\la \delta E_{\perp}^2\ra/\la \delta B_{\perp}^2\ra\sim(v_{Ai}/c)^2$, given that $v_{Ai}\propto |\bavg|/\sqrt{n}\propto \qt$, as a result of compression.

During both the exponential and secular phases of growth, the spectrum of the ion cyclotron mode is nearly monochromatic, with characteristic wavevector $\sim k_i$ in \eq{ki} and oscillation frequency $\sim\omega_i$ in \eq{omegai} --- even though, strictly speaking, Eqs.~\eqn{ki} and \eqn{omegai} only apply when the ion anisotropy is at marginal stability. At $q\,t\simeq0.55$ (vertical dashed black line in \fig{fluid1d}), the ion anisotropy is $A_i\simeq 0.5$, giving a characteristic oscillation half-period $\pi/\omega_i\simeq 12\, \omega_{0ci}^{-1}\simeq 0.12\, q^{-1}$, where we have employed the value $\omega_{0ci}/q=100$ appropriate for \fig{fluid1d}. This is comparable to the oscillation period of $\sim 0.1\, q^{-1}$ measured at $q\,t\simeq 0.55$  in \fig{fluid1d}(d) as the timespan between two consecutive minima or maxima in the temporal evolutions of $\la \delta B_y^2\ra$ or $\la \delta B_z^2\ra$, shown respectively in blue and red. We also notice that the temporal oscillations in $\la \delta E_y^2\ra$ (respectively, $\la \delta E_z^2\ra$) shown in \fig{fluid1d}(e) with a blue (respectively, red) line are correlated in time with the oscillations in  $\la \delta B_y^2\ra$ (respectively, $\la \delta B_z^2\ra$).

For the same anisotropy $A_i\simeq 0.5$, the dominant wavelength should be $\lambda_i=2\pi/k_i\simeq 2.3\,r_{L,i}$, where we have used \eq{ki} together with $r_{L,i}=\sqrt{3\beta_{0i}/2}\, c/\omega_{0\rm pi}$, for $\beta_{0i}=20$. Once again, this is in good agreement with the wavelength measured from the spatial profiles of $\delta B_y$ (solid blue) or $\delta B_z$ (solid red) in \fig{fluid1d}(f). We see that the combination of the two oppositely-propagating ion cyclotron waves (one along and one against the field, see the two signs in \eq{omegai}) results in a non-propagating pattern, with co-spatial peaks in $\delta B_y$ (solid blue) and $\delta B_z$ (solid red), with an offset of $\sim\lambda_i/2$ with respect to the peaks in $\delta E_y$ (dashed blue) and $\delta E_z$ (dashed red). In other words, if $\delta B_y, \, \delta B_z \propto \cos k_i x$, then $\delta E_y, \, \delta E_z \propto \sin k_i x$, as indeed expected from Maxwell's equations (see \eq{basic1}) and in agreement with earlier hybrid simulations \citep[e.g.,][]{mckean_92,mckean_94}. The physics of ion cyclotron waves that we have detailed so far will be used in \sect{heating} to assess the mechanism of electron heating.

For the case explored in \fig{fluid1d}, the efficiency of electron heating is quantified in \fig{fluid1d}(c), where we plot the temporal evolution of the temperature ratio $T_e/T_i=\beta_e/\beta_i$ (solid black line), its variation with respect to the initial value (blue line, dotted because the variation is negative) and the parameter $\chi-\chi_0$, which will be described below (red curve; solid if positive, dotted if negative). 

From the black and blue curves in \fig{fluid1d}(c), we see that the ratio $T_e/T_i$ steadily decreases with time, with the exception of a small glitch at $q\,t\simeq 0.55$ (see the blue dotted line). The overall trend in $T_e/T_i$ is a mere consequence of the compression of the system, and the fact that electrons are more relativistic than ions, for the parameters adopted in \fig{fluid1d} (see the electron and ion momenta in \fig{fluid1d}(b)). As we show in \app{method}, in a compressing box the particle parallel momentum does not change, whereas the perpendicular momentum increases as $p_{\perp}\propto (1+q\,t)$, so the total momentum $p=(p_\parallel^2+2\,p_{\perp}^2)^{1/2}$ increases as $p=p_0\sqrt{[1+2\,\qt^2]/3}$. In the extreme limit of ultra-relativistic electrons and non-relativistic ions, $T_e\propto \la \gamma_e v_e^2\ra\propto \la \gamma_e\ra\propto \sqrt{1+2\,\qt^2}$, whereas $T_i\propto \la \gamma_i v_i^2\ra\propto \la v_i^2\ra\propto [1+2\,\qt^2]$, so that $T_e/T_i\propto [1+2\,\qt^2]^{-1/2}$, which explains the decrease observed in the black line of \fig{fluid1d}(c).

Since we are interested in the {\it net heating of electrons by the ion cyclotron instability, and not in the straightforward effect of compression}, we choose to quantify the efficiency of electron heating by defining the ratio
\be\label{eq:chi}
\chi\equiv\frac{m_i}{m_e}\frac{\la p_e^2\ra}{\la p_i^2\ra}
\ee
which remains constant before the instability grows, and is a good indicator of the electron energy change occurring as a result of the  ion cyclotron mode. The choice of the pre-factor $m_i/m_e$ is such that in the limit of non-relativistic particles, so $\la p_e^2\ra\simeq m_e^2\la v_e^2\ra$ and $\la p_i^2\ra\simeq m_i^2\la v_i^2\ra$,  the parameter $\chi$ reduces to $\chi=m_e\la v_e^2\ra/m_i\la v_i^2\ra $, i.e., to the ratio of the average kinetic energies of electrons and ions. In \fig{fluid1d}, we  plot $|\chi-\chi_0|$, i.e., the magnitude of the $\chi$ parameter relative to its initial value.

The decrease in $\chi-\chi_0$ to negative values, starting at $q\,t\simeq 0.25$ (red dotted line in \fig{fluid1d}(c)), results from the development of the electron whistler instability, where the free energy associated with the electron anisotropy is converted into magnetic and electric energy. 
On the other hand, the development of the ion cyclotron instability at $q\,t\simeq 0.5$ results in strong electron heating (see the solid red line in \fig{fluid1d}(c)). The energy gain $\dg_{e,q}m_e c^2$, which  will be quantified in \sect{heating}, gives a corresponding change in the $\chi$ parameter of
\be\label{eq:delchi}
\Delta\chi\simeq\frac{m_e}{m_i}\frac{\la \gamma_e \dg_{e,q}\ra c^2}{\la \gamma_i^2 v_i^2\ra}~~~,
\ee
where we neglect higher order terms in $\dg_{e,q}/\gamma_e\ll1$, for reasons explained in \sect{theory}.


\section{The Physics of Electron Heating}\label{sec:heating}
In this section, we describe the physics of electron heating during the growth of the ion cyclotron instability. In \sect{theory}, we present an analytical model for the mechanism of electron heating during the development of the instability, emphasizing the physical origin of the various heating terms. In \sect{sims}, we discuss the physics of electron heating in our simulations, validating our analytical model in the case of cold and warm electrons, respectively in \sect{cold} and \sect{warm}. 

\subsection{Analytical Model of Electron Heating}\label{sec:theory}
\subsubsection{General Formalism}
We evaluate the efficiency of electron heating during the growth of the ion cyclotron instability. We employ the adiabatic approximation to describe the electron motion, since the  electron Larmor radius is much smaller than the characteristic size of the system, which is set by the wavelength $\lambda_i\sim r_{L,i}$ of the ion cyclotron mode.\footnote{For non-relativistic electrons, the requirement $r_{L,e}\ll r_{L,i}$ corresponds to $\beta_{e}/\beta_{i}\ll m_i/m_e$. For ultra-relativistic electrons, to  $\beta_{e}^2\ll \beta_{i} \,(c/v_{Ai})^2$. In both regimes, it is easily satisfied.} In the guiding center limit, the energy change of an electron --- averaged over a gyration period ---  can be written as the sum of several terms (see, e.g., \citealt{northrop_63})
\be\label{eq:en0}
\!\!\!\!\!\!\!\!\!\!\frac{\ud (\gamma-1)m_e c^2}{\ud t}&=&-e\, v_{\parallel} E_{\parallel} +\gamma\, v_{\parallel} m_e\, \bmath{v}_{E} \cdot \frac{\ud \bmath{\hat{b}}}{\ud t}+\nonumber\\
&&+\frac{\mu}{\gamma}\frac{\partial B}{\partial t}+\frac{\mu}{\gamma} \bmath{v}_{E}\!\cdot\! \nabla B+m_e\bmath{v}_{E}\!\cdot\! \frac{\ud \bmath{v}_{E}}{\ud t}
\ee
where the electromagnetic fields $\bmath{E}$ and $\bmath{B}$ are evaluated at the instantaneous location of the particle guiding center. Here, $B$ is the field magnitude, $\bmath{\hat{b}}=\bvec/B$ is the unit vector along the direction of the field and $\bmath{v}_{E}=c\,\evec\cross\bvec/B^2$ is the E-cross-B drift velocity. The Lagrangian derivative $\ud/\ud t$ (as opposed to the Eulerian derivative $\partial/\partial t$) is measured along the electron trajectory. In the guiding center approximation, the two derivatives are related by
\be\label{eq:lag}
\frac{\ud }{\ud t}=\frac{\partial}{\partial t}+(v_{\parallel}\bhat+\bmath{v}_{E})\cdot \grad~~~.
\ee

In \eq{en0}, ``parallel'' and ``perpendicular'' refer to the direction of the local magnetic field (so, $E_\parallel=\evec\cdot\bhat$). The electron, of mass $m_e$, charge $-e$ and Lorentz factor $\gamma$, has a parallel velocity $v_{\parallel}$ and a perpendicular velocity $\sqrt{2}\,v_{\perp}$, so that its total velocity is $v=(v_\parallel^2+2 v_\perp^2)^{1/2}$.\footnote{In this section, for ease of notation we remove the subscript ``e'' to indicate electrons.} Note that our choice for the definition of the perpendicular velocity is such that for an isotropic distribution $\langle v_\parallel^2\rangle=\langle v_\perp^2\rangle$. The electron adiabatic moment is $\mu=\gamma^2 v_\perp^2m_e/B $. If the  magnetic fields generated by the ion cyclotron instability are weaker than the background ordered field (i.e., $\dbperp/|\bavg|\ll1$), we can simply assume that $v_\parallel=v_x$ and $v_\perp=[(v_y^2+v_z^2)/2]^{1/2}$, since in our configuration the ordered magnetic field lies along $x$. 

In \eq{en0}, the first line corresponds to energy gain (or loss) involving the parallel energy. The first term describes  linear acceleration by the component of electric field aligned with the magnetic field, whereas the second term includes (but is not limited to) Fermi-like acceleration induced by the curvature of the field lines. More precisely, using \eq{lag}, the second term in the first line of \eq{en0} can be recast as
\be\label{eq:en1}
\gamma\, v_{\parallel} m_e\, \bmath{v}_{E} \!\cdot\! \frac{\ud \bmath{\hat{b}}}{\ud t}&=&\gamma\, v_{\parallel} m_e\, \bmath{v}_{E}\!\cdot\! \frac{\partial \bhat}{\partial t}+\gamma\, v_{\parallel} m_e\, \bmath{v}_{E}\!\cdot\![(\bmath{v}_{E}\!\cdot\! \grad)\bhat]+\nonumber\\&&+\,\gamma\, v_{\parallel}^2 m_e\bmath{v}_{E}\cdot[(\bhat\cdot \grad)\bhat]~~~.
\ee
The last term in \eq{en1} describes the so-called curvature drift. In fact, it is straightforward to rewrite this term as $-e\, \bmath{v}_{\rm curv}\cdot \evec$, where the curvature velocity is
\be
\bmath{v}_{\rm curv}=-\frac{\gamma v_{\parallel}^2m_e c}{eB} \bhat\cross[(\bhat\cdot \grad)\bhat]~~~.
\ee

In \eq{en0}, the second line represents the energy gains and losses involving the perpendicular energy. The first term in this line is the induction effect, caused by the curl of the electric field acting around the circle of gyration. The second term represents the energy gain associated with the grad-B drift. In fact, it can be rewritten as $-e\, \bmath{v}_{\rm \grad B}\cdot \evec$, where the grad-B speed is 
\be
\bmath{v}_{\grad B}=-\frac{\gamma v_{\perp}^2m_e c}{eB}\frac{\bhat\cross\nabla B}{B}~~~.
\ee
Finally, the third term can be further separated into three contributions. Using \eq{lag}, we obtain
\be\label{eq:pol}
\!\!\!\!\!\!\!\!\!\!m_e\bmath{v}_{E}\!\cdot\! \frac{\ud \bmath{v}_{E}}{\ud t}\!=\!m_e\bmath{v}_{E}\!\cdot\! \frac{\partial \bmath{v}_{E}}{\partial t}\!+\!m_e\bmath{v}_{E}\!\cdot\![(v_{\parallel}\bhat+\bmath{v}_{E})\!\cdot\! \grad]\bmath{v}_{E}
\ee
The first term on the right hand side of \eq{pol} can be interpreted as the energy gain associated with the polarization drift. It can be recast as $-e\, \bmath{v}_{\rm pol}\cdot \evec$, where the polarization velocity is defined as
\be
\bmath{v}_{\rm pol}=-\frac{m_e}{eB}\bhat\cross\frac{\partial \bmath{v}_E}{\partial t}~~~.
\ee


\subsubsection{Electron Heating during the Ion Cyclotron Instability}\label{sec:terms}
We now proceed to estimate the contribution of the different terms in the case of electron heating by the ion cyclotron instability. By averaging over the electron population, we can assume that $\langle v_\parallel\rangle=0$ and thus neglect all the terms linear in $v_\parallel$. The mean electron energy gain then turns out to be\footnote{Unless  noted otherwise, the symbol  $\la Q \ra$ should be interpreted as an average over the local particle distribution if the quantity $Q$ pertains to electrons (as in this equation), or as an average over space if the quantity $Q$ refers to electromagnetic fields (as in some later equations).}
\be\label{eq:en2}
\!\!\!\!\!\frac{\ud \la\gamma-1\ra m_e c^2}{\ud t}\!&=&\langle\gamma\, v_{\parallel}^2\rangle\, m_e\bmath{v}_{E}\cdot[(\bhat\cdot \grad)\bhat]+\\
&&+\frac{\langle\gamma v_{\perp}^2\rangle m_e}{B}\!\left(\frac{\partial B}{\partial t}\!+ \!\bmath{v}_{E}\!\cdot\! \nabla B\!\right)\!\!+\!\frac{m_e}{2} \frac{\partial \bmath{v}_{E}^2}{\partial t}~,\nonumber
\ee
where in the last term we have replaced the Lagrangian derivative with the Eulerian derivative, neglecting terms of order $|\bmath{v}_{E}|^3/c^3\ll1$ in \eq{pol}. In addition, while the electromagnetic fields in \eq{en0} had to be measured at the guiding center of the selected particle, in \eq{en2} we have implicitly assumed that the fields can be evaluated at the particle instantaneous location (in other words, that the electromagnetic fields have no significant correlation with the properties of single electrons). Below, we explicitly show that this is a good approximation. 

In order to estimate the different heating contributions during the phase of exponential growth of the ion cyclotron instability, we need to prescribe the evolution of the electromagnetic fields. As we have discussed in \sect{struct}, the spectrum of unstable modes is nearly monochromatic, with characteristic wavevector $k_i$ (\eq{ki}) and frequency $\omega_i\simeq k_i v_{Ai}$ (\eq{omegai}). Based on the findings in \sect{struct1}, we can write the magnetic field components as 
\be\label{eq:bfield}
\delta B_y&=&C_B(t) \cos\omega_i t\, \cos k_i x~~,\nonumber\\
\delta B_z&=&C_B(t) \sin\omega_i t\, \cos k_i x~~,
\ee
where we have assumed that the wavevector is aligned with the $x$ direction, as expected in our setup. The function $C_B(t)$ describes the exponential growth of the fields during the ion cyclotron instability. The exponential phase lasts for a time $\sim0.1-0.2\, q^{-1}$, which is longer than the oscillation period $\sim \omega_i^{-1}\sim \omega_{ci}^{-1}$ (assuming $\omega_{ci}/q\gg1$, as expected in accretion flows), yet shorter than the characteristic compression timescale $q^{-1}$. Therefore, when focusing on the exponential phase of the ion cyclotron mode, we neglect the overall compression of the box (i.e., we assume that $(1+q\,t)$ is nearly constant, and in particular that the mean field $|\bavg|$ does not significantly change over time).

From Maxwell's equations, one can obtain from \eq{bfield} the temporal evolution of the perpendicular components of the electric field, 
\be\label{eq:efield1}
\delta E_y&=&-\frac{1}{k_ic}[\dot{C}_B \sin\omega_i t+C_B \omega_i \cos \omega_i t] \sin k_i x~~,\nonumber\\
\delta E_z&=&\frac{1}{k_ic}[\dot{C}_B \cos\omega_i t-C_B \omega_i \sin \omega_i t] \sin k_i x~~,
\ee
where $\dot{C}_B=\ud C_B/\ud t$. With the expressions in \eq{bfield} and \eq{efield1} for the electromagnetic fields of the growing ion cyclotron instability, we can now evaluate the time-integrated contributions of the terms in \eq{en2}. We take the limit  $C_B/|\bavg|\sim\dbperp/|\bavg|\ll1$ of weak turbulence, so that the total magnetic field can be expanded as $B\simeq |\bavg|+ \delta B_{\perp}^2/2|\bavg|$. Moreover, the E-cross-B drift and the magnetic unit vector are, to leading order,
\be
\!\!\!\!\frac{\bmath{v}_{E}}{c}&=&\left\{\frac{\delta E_y \delta B_z-\delta E_z \delta B_y}{|\bavg|^2},\frac{\delta E_z}{|\bavg|},-\frac{\delta E_y}{|\bavg|}\right\}~~,\\ 
\bhat&=&\left\{1,\frac{\delta B_y}{|\bavg|},\frac{\delta B_z}{|\bavg|}\right\}~~~.
\ee

As we have discussed above, the first term on the right hand side of \eq{en2} corresponds to the energy change associated with the curvature drift, so its time-integrated contribution will be called $\Delta \gamma_{\rm curv}m_e c^2$. To leading order in $\dbperp/|\bavg|\ll1$, this curvature term equals
\be
\frac{\langle\gamma\, v_{\parallel}^2\rangle\, m_ec}{|\bavg|^2}\left[\delta E_z\frac{\partial (\delta B_y)}{\partial x}-\delta E_y \frac{\partial (\delta B_z)}{\partial x}\right]=\nonumber\\-\langle\gamma\, v_{\parallel}^2\rangle\, m_e\frac{\dot{C}_B C_B}{|\bavg|^2}\sin^2k_i x~~~.
\ee
By integrating over time, under the assumption that both $\la \gamma v_{\parallel}^2\ra$ and $|\bavg|$ do not appreciably change during the exponential phase of the instability (as we  will justify a posteriori), we find that the time-integrated energy change associated with the curvature drift is
\be\label{eq:dgcurva}
\Delta\gamma_{\rm curv}m_e c^2\simeq -\left[\la\gamma v_{\parallel}^2\ra m_e\frac{C_B^2\sin^2\!k_i x}{2|\la \bvec \ra|^2}\right]_{\rm exp}~~,
\ee
where the subscript ``exp'' prescribes that the quantity in parentheses should be evaluated at the end of the phase of exponential growth.
The negative sign in \eq{dgcurva} indicates that the curvature term results in cooling, rather than heating, and it is largest at the locations where $\delta E_{\perp}^2\propto \sin^2 k_i x$ peaks, as we demonstrate in \sect{sims}. Since the fractional energy loss is of order $\sim C_B^2/|\la \bvec \ra|^2\ll1$, it follows that $\langle \gamma v_{\parallel}^2\ra$  is nearly constant during the growth of the instability, as we had assumed. 

The time-integrated contribution of the term $\langle\gamma v_{\perp}^2\rangle m_eB^{-1}\partial B/\partial t$ in the second line of \eq{en2}, which we shall call $\Delta\gamma_{\partial B}m_e c^2$ given that it depends on the temporal derivative of the magnetic field, is easy to estimate. In the limit $\dbperp/|\bavg|\ll1$ of weak turbulence, we have $B\simeq |\bavg|+ \delta B_{\perp}^2/2|\bavg|$. The time derivative $\partial|\bavg|/\partial t $ simply accounts for compression-induced heating. Since this is a mere consequence of our computational setup --- occurring even without any instability --- it will be neglected hereafter. 
To leading order, the first non-trivial contribution to electron heating is provided by the time derivative of $ \delta B_{\perp}^2/2|\bavg|$. This vanishes in the phase of secular growth of the ion cyclotron instability, since both $\delta B_{\perp}^2$ and the mean field $|\bavg|$ grow as $(1+q\,t)^2$, as we have discussed in \sect{struct1}. It follows that significant electron heating only occurs during the exponential phase of the ion cyclotron mode. Assuming that the factor $\langle \gamma v_{\perp}^2\rangle/B\sim \langle \gamma v_{\perp}^2\rangle/|\la \bvec \ra|$ in \eq{en2} is nearly constant in time (this will be justified a posteriori; notice that, in the limit of non-relativistic electrons, it is equivalent to the conservation of the magnetic moment), the time integration yields
\be\label{eq:dgpartial}
\Delta\gamma_{\partial B}m_e c^2\simeq \left[\la\gamma v_{\perp}^2\ra m_e\frac{C_B^2\cos^2\!k_i x}{2|\la \bvec \ra|^2}\right]_{\rm exp}~~.
\ee
This heating term is largest at the locations where $\delta B_{\perp}^2\propto \cos^2 k_i x$ peaks, as we demonstrate in \sect{sims}. The fractional energy gain is of order $\sim \delta B_{\perp}^2/|\la \bvec \ra|^2\ll1$, so that $\langle \gamma v_{\perp}^2\ra$  is nearly constant in time, which justifies our initial assumption.

The second term in the second line of \eq{en2} describes the effect of the grad-B drift, and its time-integrated contribution will be indicated as $\Delta \gamma_{\grad B}m_e c^2$. For the magnetic field in \eq{bfield}, it reduces to
\be
\!\!\!\frac{\langle\gamma\, v_{\perp}^2\rangle\, m_e}{2|\bavg|^2}(\bmath{v}_E\cdot \bmath{\hat{x}})\frac{\partial (\delta B_\perp^2)}{\partial x}=\langle\gamma\, v_{\perp}^2\rangle\, m_e\frac{\dot{C}_B C_B^3}{|\bavg|^4}~~.
\ee
As before, we assume that both $\la \gamma v_{\perp}^2\ra$ and $|\bavg|$ do not appreciably change during the exponential phase of the instability. By integrating over time, we find 
\be\label{eq:dggradb}
\!\!\!\!\!\Delta \gamma_{\grad B}m_e c^2\simeq  \left[\la\gamma v_{\perp}^2\ra m_e\frac{C_B^4\sin^2\!k_i x\,\cos^2\!k_i x}{4|\la \bvec \ra|^4}\right]_{\rm exp}~,
\ee
which is smaller than the contribution in \eq{dgpartial} by a factor of $\dbperp^2/|\bavg|^2\ll1$.\footnote{This term is neglected in second-order theories of electron heating, see \citet{gary_feldman_78,gary_93}. However, as we show in \eq{dgnet}, the sum of $\Delta \gamma_{\rm curv} $ and $\Delta \gamma_{\partial B}$ tends to vanish in high-beta plasmas, where the electron anisotropy stays small. In this case, the $\Delta \gamma_{\grad B}$ term in \eq{dggradb} might dominate the electron energy gain.} Once again, the spatial dependence of this term will be confirmed in \sect{sims}.

Finally, the last term in \eq{en2} gives the energy gain associated with the E-cross-B velocity. To leading order, we find that $\bmath{v}^2_E\simeq c^2\delta E_{\perp}^2/ |\bavg|^2$. Since both $\delta E_{\perp}^2$ and $ |\bavg|^2$ scale as $(1+q\,t)^4$ during the secular phase of growth, as we have demonstrated in \sect{struct1}, the time derivative in \eq{en2} is nonzero only during the exponential phase of the ion cyclotron instability. At the end of this phase, the resulting energy gain is 
\be\label{eq:dgexb}
\Delta \gamma_{E\cross B}m_e c^2\simeq \left[m_e\left(\frac{\omega_i}{k_i}\right)^2\frac{C_B^2\sin^2\!k_i x}{2 |\bavg|^2}\right]_{\rm exp}~~,
\ee
where $\omega_i/k_i\simeq v_{Ai}$ is the phase speed of the ion cyclotron waves. We remark that, unlike the terms we have previously discussed, the contribution in \eq{dgexb} is independent of the initial electron energy. Even for electrons that start with extremely low temperatures, the term in \eq{dgexb} provides a solid temperature ``floor''. By comparing \eq{dgexb} with \eq{dgpartial}, we argue that the electron energy gain will be dominated by the E-cross-B contribution if the initial electron temperature satisfies $k_B T_{e}\lesssim m_e v_{Ai}^2$, or equivalently $\beta_{e}\lesssim 2 \,m_e/m_i$.

In summary, the average increase in the electron Lorentz factor resulting from the growth of the ion cyclotron instability can be written as
\be\label{eq:enall}
\!\!\!\!\!\!\!\dg_{e,q}\equiv\gamma_e - \gamma_{e,q}&=&-|\Delta \gamma_{\rm curv}|+\nonumber\\&&+\Delta \gamma_{\partial B}\!+\!\Delta \gamma_{\grad B}\!+\!\Delta \gamma_{E\cross B}~,
\ee
where we have subtracted the term $\gamma_{e,q}$ on the left hand side, since it accounts for the electron energy increase due to compression alone, which would be present even without any instability. Finally, we remark that $\Delta \gamma_{\rm curv} $ and $\Delta \gamma_{\partial B}$ have the same scaling with the turbulent field $\sim C_B^2/|\bavg|^2$, but opposite sign. Once averaged in space, their sum equals
\be\label{eq:dgnet}
\!\!\langle\Delta \gamma_{\partial B}+\Delta \gamma_{\rm curv}\rangle_{}\!\simeq\! \left[\frac{\la\gamma v_{\perp}^2\ra\!-\!\la\gamma v_{\parallel}^2\ra}{c^2}\,\frac{C_B^2}{4|\la \bvec \ra|^2}\right]_{\rm exp}~,
\ee
where the average $\la\Delta \gamma_{\partial B}+\Delta \gamma_{\rm curv} \ra_{}$ is over all the electrons in the system. From \eq{dgnet}, we conclude that the net contribution only depends on the anisotropy of the electron population, and it vanishes for isotropic electrons, since $\la \gamma v_{\perp}^2\ra=\la \gamma v_{\parallel}^2\ra$. Also, we remark that, due to the cancellation between the two terms in \eq{dgnet}, the criterion for the dominance of the term in \eq{dgnet} over the E-cross-B contribution in \eq{dgexb} should be properly written as $\beta_{e} A_e\gtrsim 2 \,m_e/m_i$, where $A_e$ is the electron anisotropy. For simplicity, in our discussion below we shall assume $A_e\sim 1$ (see the solid blue line in \fig{fluid1d}(d)), so that the criterion is simply $\beta_{e}\gtrsim 2 \,m_e/m_i$.

\subsection{Electron Heating in PIC Simulations}\label{sec:sims}
\begin{figure*}[tbp]
\begin{center}
\includegraphics[width=0.77\textwidth]{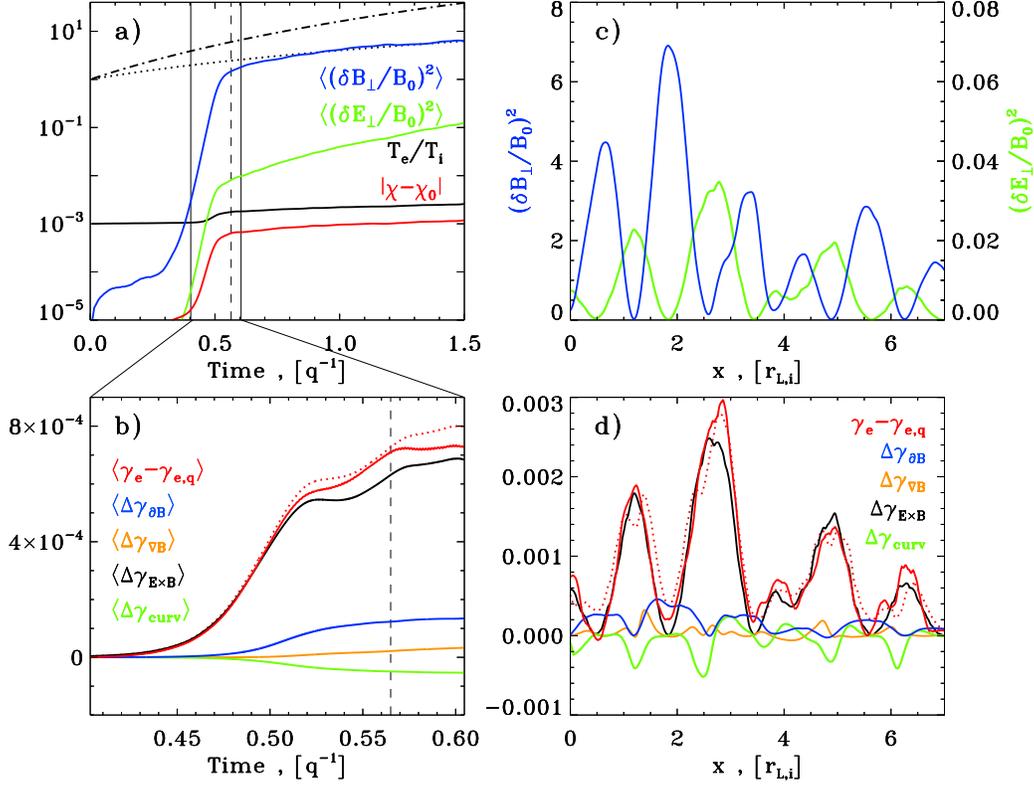}
\caption{Physics of electron heating, from a representative 1D simulation with cold electrons, i.e., $\beta_{0e}\lesssim 2\, m_e/m_i$. We choose $\beta_{0i}=20$, $T_{0e}/T_{0i}=\ex{3}$, $v_{A0i}/c=0.05$, $\omega_{0ci}/q=100$ and $m_i/m_e=16$. We plot: (a) The overall temporal evolution of the system, showing the development of the ion cyclotron instability, which results in magnetic and electric energy (blue and green line, respectively), and in electron heating, as quantified by the ratio $T_e/T_i$ of the electron to ion temperature, or by the parameter $|\chi-\chi_0|$, where $\chi$ is defined in \eq{chi} and $\chi_0$ is its initial value. For reference, we also plot $\qt^2$ (dotted black line) and $\qt^4$ (dot-dashed black line). In the timespan delimited by the vertical solid black lines in panel (a), we show in panel (b) the various contributions to electron heating (see legend). The sum of the different contributions (dotted red line), which is dominated by  the E-cross-B term (solid black line), follows closely the actual evolution measured from the simulation (solid red line), validating our analytical model. We then localize the source of heating in space, by comparison with the spatial profiles of the magnetic (blue) and electric (green) energy densities shown in panel (c), as a function of the longitudinal coordinate $x$, measured in units of the ion Larmor radius $r_{L,i}$. As presented in panel (d), the various heating contributions (see legend) have different spatial profiles. The plot of the actual electron energy gain (solid red), follows closely the sum of the different contributions (dotted red), which is dominated by the E-cross-B term (solid black line).
}
\label{fig:cold}
\end{center}
\end{figure*}

\begin{figure*}[!htbp]
\begin{center}
\includegraphics[width=0.77\textwidth]{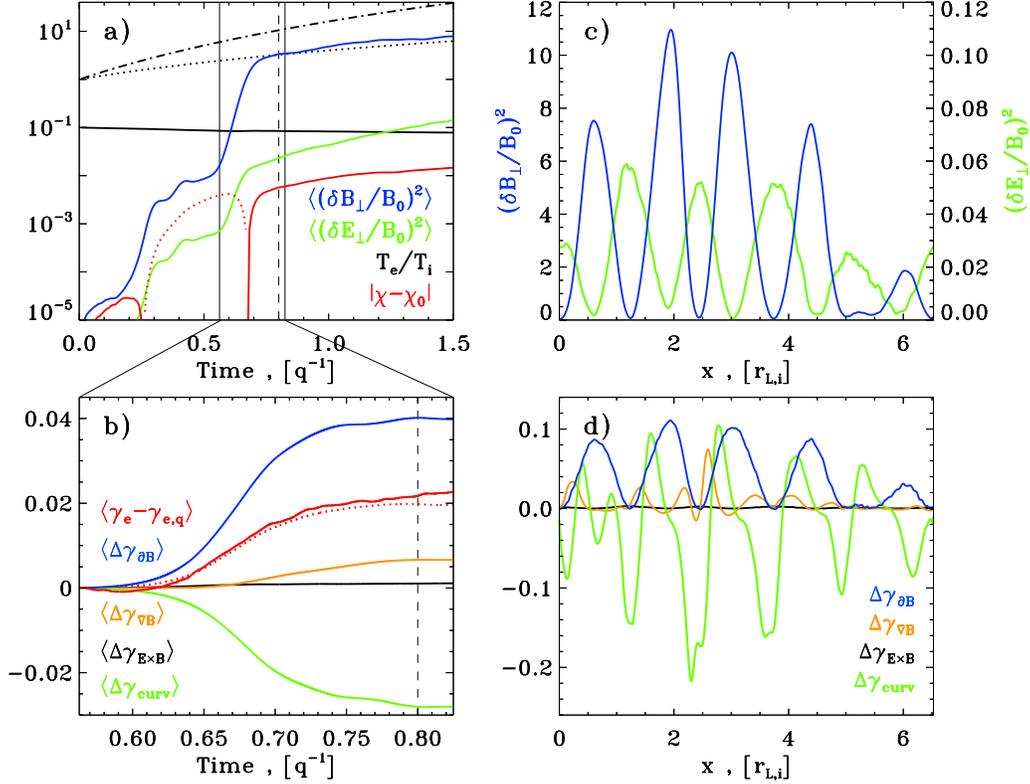}
\caption{Physics of electron heating, from a representative 1D simulation with warm electrons, i.e., $\beta_{0e}\gtrsim 2\, m_e/m_i$. We choose $\beta_{0i}=20$, $T_{0e}/T_{0i}=\ex{1}$, $v_{A0i}/c=0.05$, $\omega_{0ci}/q=50$ and $m_i/m_e=64$. The panels show the same quantities as in \fig{cold}, with the following exceptions. The dotted red line in panel (a) is chosen to indicate when the heating parameter $\chi-\chi_0$ is negative (instead, solid line for positive values). In panel (d), we do not show the spatial profile of the electron energy gain (red lines in \fig{cold}(d)), since the heating process is non-local, i.e., the electrons move across several wavelengths during the growth of the ion cyclotron mode. For warm electrons, the energy change is dominated by the balance between the gain due to $\Delta \gamma_{\partial B}$  and the loss due to $\Delta \gamma_{\rm curv}$, as shown in panels (b) and (d).
}
\label{fig:warm}
\end{center}
\end{figure*}

In this section, we demonstrate that the analytic model described in \sect{theory} is in excellent agreement with the results of our simulations. In \fig{cold} and \fig{warm}, we analyze in detail the various contributions to electron heating during the growth of the ion cyclotron mode, for the cases of cold ($\beta_{0e}\lesssim 2\, m_e/m_i$) and warm ($\beta_{0e}\gtrsim 2\, m_e/m_i$) electrons, respectively. In both cases, we show in panel (a) the complete temporal evolution of the transverse magnetic and electric energy, of the electron-to-ion temperature ratio and of the $\chi$ parameter defined in \eq{chi}. Then, we focus on the exponential growth of the ion cyclotron mode (as delimited by the vertical solid black lines in panel (a)), and present in panel (b)   the resulting increase in the mean electron Lorentz factor, and the partition of such energy gain into the various terms described in \sect{theory}. In panel (c) we show the spatial profiles of magnetic and electric energy at the saturation of the ion cyclotron growth, and in panel (d) the spatial dependence of the different heating terms, at the same time as in panel (c) (the selected time is marked with the  vertical dashed black line in panels (a) and (b)).

For a complete characterization of the electron energy gain in the adiabatic approximation, one should use \eq{en0}, calculating for each electron the electromagnetic fields and their derivatives at the instantaneous location of the particle gyrocenter. We have performed such an experiment, following a sample of $\sim 10,000$ electrons, whose instantaneous properties (location, momentum and local electromagnetic fields) were saved every $10$ timesteps. For each electron, we  computed all the terms on the right hand side of \eq{en0} and confirmed that the mean electron energy gain (which can be calculated directly from the simulation, from the temporal evolution of the particle Lorentz factor) is equal to the time-integrated sum of the various contributions on the right hand side of \eq{en0}. So, \eq{en0} properly describes the physics of electron heating.

We then proceeded one step further, checking that the assumptions leading to \eq{en2} are satisfied. We verified that direct acceleration by the parallel electric field (the first term on the right hand side of \eq{en0}) is always negligible, once averaged over the electron population. We also checked that the only significant contribution in \eq{en1} is provided by the proper curvature term, i.e., the one proportional to $v_\parallel^2$. We have tested that the electromagnetic fields can be  measured either at the instantaneous particle location (as in \eq{en2}) or at the particle guiding center (as in \eq{en0}), without appreciable differences, once we average over the electron population. Finally, we checked that $\la \partial \bmath{v}_E^2/\partial t\ra\simeq\la \ud \bmath{v}_E^2/\ud t\ra$. 

In summary, we have used a sample of particles extracted from our simulations to explicitly check that all the assumptions leading to \eq{en2} are well motivated. Below, we integrate over time the terms on the right hand side of \eq{en2}.
The resulting space-dependent contributions shall be called respectively $\Delta \gamma_{\rm curv}$, $\Delta \gamma_{\partial B}$, $\Delta \gamma_{\grad B}$ and $ \Delta \gamma_{E\cross B}$, and their spatial averages will be $\la\Delta \gamma_{\rm curv}\ra$, $\la\Delta \gamma_{\partial B}\ra$, $\la\Delta \gamma_{\grad B}\ra$ and $\la \Delta \gamma_{E\cross B}\ra$. Note that we use the same notations as in \sect{terms} (see Eqs.~\eqn{dgcurva}, \eqn{dgpartial}, \eqn{dggradb} and \eqn{dgexb}). However, in the two subsections below, we directly integrate \eq{en2}, without making use of the approximations employed in  \sect{terms} to derive the simple expressions in Eqs.~\eqn{dgcurva}, \eqn{dgpartial}, \eqn{dggradb} and \eqn{dgexb}. Yet, we have explicitly checked that the assumptions leading to these equations are realized in our simulations.


In the two subsections below, we present results for two representative cases, one with cold electrons ($\beta_{0e}\lesssim 2\, m_e/m_i$) and one with  warm electrons ($\beta_{0e}\gtrsim 2\, m_e/m_i$). We have checked the consistency of our results --- by following in each case a sample of $\sim 10,000$ electrons from the simulation --- for three values of ion beta ($\beta_{0i}=5,\,20$ and 80), two values of the electron-to-ion temperature ratio ($T_{0e}/T_{0i}=\ex{3}$ and $\ex{1}$), two choices of the Alfv\'en velocity ($v_{A0i}/c=0.05$ and 0.1) and of the ratio $\omega_{0ci}/q$ ($=50$ and 100) and two values of the mass ratio ($m_i/m_e=16$ and 64). In all cases, our results are in very good agreement with the model presented above. More details on the dependence of the heating efficiency on flow conditions will be presented in Paper II.


\subsubsection{Cold Electrons: $\beta_{0e}\lesssim 2\,m_e/m_i$}\label{sec:cold}
In \fig{cold}, we present the physics of electron heating by the ion cyclotron instability, in the case  of cold electrons: $\beta_{0e}\lesssim 2 \,m_e/m_i$. We choose $\beta_{0i}=20$, $T_{0e}/T_{0i}=\ex{3}$, $v_{A0i}/c=0.05$, $\omega_{0ci}/q=100$ and $m_i/m_e=16$. It follows that $\beta_{0e}=0.02\ll m_e/m_i$.

The development of the ion cyclotron instability, as shown by the temporal evolution of the transverse magnetic energy $\la \delta B_\perp^2\ra$ and of the transverse electric energy $\la \delta E_{\perp}^2\ra$ (blue and green lines in \fig{cold}(a), respectively) follows closely the evolution shown in \fig{fluid1d}, despite the difference in $T_{0e}/T_{0i}$ and in $m_i/m_e$. In particular, in both cases the magnetic energy reaches $\la \delta B_{\perp}^2\ra/|\bavg|^2\sim0.3$ at the end of the exponential phase ($q\,t\simeq0.55$), whereas the electric energy at the same time is smaller by a factor of $\sim (v_{A0i}/c)^2\simeq2.5\times\ex{3}$. So, the development of the ion cyclotron instability does not depend on the electron thermal content or the ion-to-electron mass ratio.

As a result of the growth of the ion cyclotron instability, the electron-to-ion temperature ratio increases up to $T_e/T_i\simeq 2\times \ex{3}$ (black line in \fig{cold}(a)). Accordingly, the $\chi$ parameter grows exponentially around $q\,t\simeq0.5$, saturating at $\chi-\chi_0\simeq \ex{3}$ (red line in \fig{cold}(a)). The net increase in the mean electron Lorentz factor --- after excluding the effect of compression --- is shown in \fig{cold}(b) with a solid red line. We focus on the exponential growth phase of the ion cyclotron mode, i.e., $0.4\lesssim q\,t\lesssim0.6$ (as marked by the vertical solid black lines in \fig{cold}(a)). As expected for $\beta_{0e}\lesssim 2 \,m_e/m_i$ (cold electrons), most of the electron energy gain is provided by the E-cross-B term $\la \Delta \gamma_{E\cross B}\ra$ (black line in \fig{cold}(b)). The other three terms on the right hand side of \eq{en2}, which depend on the initial electron temperature, have only a minor contribution, with $\la \Delta \gamma_{\partial B}\ra$ and $\la \Delta \gamma_{\rm curv}\ra$ having opposite signs (blue and green in \fig{cold}(b), respectively), in agreement with \eq{dgnet}. The sum of the four contributions (dotted red line in \fig{cold}(b)) follows closely the net increase in Lorentz factor as measured directly from the simulation (solid red line), confirming that \eq{en2} can be confidently used to estimate the efficiency of electron heating.

We remark that, since the E-cross-B term $\la \Delta \gamma_{E\cross B}\ra$ does not depend on the initial electron energy (see the last term in \eq{en2}, which results in \eq{dgexb}), it provides a solid temperature ``floor'' even for electrons starting with virtually zero temperature. On the other hand, the magnitude of this term depends explicitly on the electron-to-ion mass ratio, so that the resulting increase in the $\chi$ parameter will scale as $\propto m_e/m_i$, everything else being fixed. We have directly verified that, at fixed $\beta_{0e}m_i/m_e\ll1$, the electron heating efficiency is dominated by the E-cross-B term and so it scales in inverse proportion with the ion-to-electron mass ratio (we have tested this for $m_i/m_e=16$ and 64).

As regards the spatial dependence of electron heating, we expect on analytical grounds that $\Delta \gamma_{E\cross B}\propto \bmath{v}^2_E \propto \delta E_{\perp}^2$. The spatial profile of $\delta E_{\perp}^2$ is shown in \fig{cold}(c) with a green line (instead, $\delta B_{\perp}^2$ is plotted in blue) at the time $q\,t=0.565$ that is marked with a vertical dashed black line in panels (a) and (b). \fig{cold}(d) shows that the time-integrated energy gain due to the E-cross-B term (black line) follows closely the spatial profile of $\delta E_\perp^2$ (green line in \fig{cold}(c)), in agreement with our analytical model. The spatial profiles of the other three terms (blue for $\Delta \gamma_{\partial B}$, green for $\Delta \gamma_{\rm curv}$ and orange for $\Delta \gamma_{\grad B}$ in \fig{cold}(d)) also agree with the model, but we postpone their analysis to the following subsection, where their role in electron heating is much more important. 

From \fig{cold}(d), it is remarkable that the spatial profile of the predicted electron energy gain (dotted red line) agrees well with the spatial dependence of the actual electron energy increase (solid red line). This suggests that, in the case of cold electrons, the final electron energy  depends dramatically on the initial spatial location, since electrons are heated locally by the growing electric fields (which, as discussed in \sect{struct1}, form a stationary pattern in space). The electron distribution at each location resembles a Maxwellian drifting with the local E-cross-B velocity. We have directly tested the local nature of the heating process in the case of cold electrons, by performing a selected run in which the position of each electron was artificially kept constant. We find that, despite this artificial constraint on the electron motion, the heating efficiency is unchanged, which confirms the local nature of the heating process.



\subsubsection{Warm Electrons: $\beta_{0e}\gtrsim 2\,m_e/m_i$}\label{sec:warm}
To study the case of warm electrons with $\beta_{0e}\gtrsim 2\,m_e/m_i$, we increase both the electron-to-ion temperature ratio to $T_{0e}/T_{0i}=\ex{1}$ and the ion-to-electron mass ratio to $m_i/m_e=64$. It follows that $\beta_{0e}=2\gg m_i/m_e$. 

The parameters chosen for \fig{warm} --- $\beta_{0i}=20$, $T_{0e}/T_{0i}=\ex{1}$, $v_{A0i}/c=0.05$, $\omega_{0ci}/q=50$ and $m_i/m_e=64$ --- are identical to the reference case discussed in \sect{struct1} (see \fig{fluid1d}), with the exception of $\omega_{0ci}/q$, which is smaller here by a factor of two. Despite this difference, we notice that the magnetic energy  at the end of the exponential phase is still $\la \delta B_{\perp}^2\ra/|\bavg|^2\sim0.3$ (blue line in \fig{warm}), and the electric energy is still  a factor of $\sim (v_{A0i}/c)^2\simeq2.5\times\ex{3}$ smaller than the magnetic energy (green line in \fig{warm}). The only variation, with respect to the case $\omega_{0ci}/q=100$ studied in \sect{struct1}, is the fact that the exponential growth terminates at $q\,t\simeq0.7$, later than in \fig{fluid1d}. 

For the parameters adopted in \fig{warm}, the electrons are quasi-relativistic. Due to the different adiabatic index with respect to the non-relativistic ions, the electrons are disfavored, as regard to compressive heating. This explains the steady decrease in $T_e/T_i$ shown in \fig{warm}(a) with a black line.  Excluding the effect of compression, the $\chi$ parameter (red line in \fig{warm}(a)) shows an early phase of cooling (dotted red line in \fig{warm}(a) at $0.25\lesssim q\,t\lesssim0.6$), due to the growth of the electron whistler instability, which drains the free energy of the electron anisotropy to generate electromagnetic waves (see the corresponding bumps in the blue and green lines in  \fig{warm}(a)). At $q\, t\gtrsim 0.6$, the ion cyclotron instability grows and heats the electrons, bringing the $\chi$ parameter back to positive values (solid red line in \fig{warm}(a)), up to a saturation value of $\chi-\chi_0\simeq\ex{2}$.

The net increase in the mean electron Lorentz factor --- excluding the effect of compression --- is shown in \fig{warm}(b) with a solid red line, during the exponential growth of the ion cyclotron mode ($0.55\lesssim q\,t\lesssim0.8$, as marked by the vertical solid black lines in \fig{warm}(a)). We notice that, following a decrease at early times which is associated with the electron whistler mode (see the minor dip in the solid red line at $q\,t\simeq 0.6$ in \fig{warm}(b)), the mean electron Lorentz factor asymptotes to $\simeq 0.02$. 

The sum of the four contributions on the right hand side of \eq{en2}, once integrated over time (dotted red line in \fig{warm}(b)), follows closely the net increase in Lorentz factor as measured directly from the simulation (solid red line), confirming that \eq{en2} (and consequently, \eq{enall}) can be confidently used to estimate the efficiency of electron heating. In this case of warm electrons, the E-cross-B term $\la \Delta \gamma_{E\cross B}\ra$ (black line in \fig{warm}(b)), which dominated for cold electrons, is completely negligible.
 Rather, most of the electron heating comes from the combination of the (positive) term $\la \Delta \gamma_{\partial B}\ra$ (blue line in \fig{warm}(b)) and the (negative) curvature term $\la \Delta \gamma_{\rm curv}\ra$ (green line), with a minor contribution by the grad-B term $\la \Delta \gamma_{\grad B}\ra$ (orange line). Unlike the E-cross-B factor, these three terms  yield an  increase in electron energy proportional to the electron temperature, irrespective of the electron-to-ion mass ratio.

By comparing \fig{warm}(b) with \fig{cold}(b), we can demonstrate that the heating contributions  $\la \Delta \gamma_{\partial B}\ra$ (blue line) and $\la \Delta \gamma_{\rm curv}\ra$ (green line) scale as predicted in Eqs.~\eqn{dgpartial} and \eqn{dgcurva}, respectively. In terms of the variation in Lorentz factor, which is plotted in panel (b), the blue and green curves should scale as $\propto (T_{0e}/T_{0i}) (m_i/m_e)$, implying a difference of a factor of $\sim 400$ between \fig{cold} and \fig{warm}, in good agreement with the numerical results. Most importantly, the sum of the four contributions (dotted red line in \fig{warm}(b)) follows closely the net increase in Lorentz factor as measured directly from the simulation (solid red line), confirming that \eq{en2} (and consequently, \eq{enall}) are valid even for warm electrons. 

To investigate the spatial dependence of the different heating contributions, we present the spatial profiles of the magnetic (blue) and electric (green) energy in \fig{warm}(c). We find that the electron temperatures perpendicular and parallel to the mean field are nearly uniform in space (not shown), with a modest tendency for lower perpendicular temperatures (and higher parallel temperatures) close to the peaks in magnetic energy, as expected from the mirror effect. The spatial uniformity of $T_\perpe$ and $T_\pare$ is related to the fact that the heating process is non-local, i.e., the electrons move across several wavelengths during the growth of the ion cyclotron mode. This is the reason why in  \fig{warm}(d) we do not show the spatial profile of the electron energy gain, as opposed to \fig{cold}(d) (red solid and dotted lines), where the heating mechanism was spatially localized.

In \fig{warm}(d), we confirm the spatial dependence of the different heating terms, as described in Eqs.~\eqn{dgcurva}, \eqn{dgpartial} and \eqn{dggradb}. In \sect{theory}, we have assumed, among other things, that the spectrum of the ion cyclotron mode is monochromatic, as in \eq{bfield}, and we have derived that  $ \Delta \gamma_{\partial B}\propto \delta B_\perp^2\propto \cos^2 \!k_i x$, $\Delta \gamma_{\rm curv}\propto \sin^2 \!k_i x$ and $ \Delta \gamma_{\grad B}\propto \cos^2\! k_i x\,\sin^2\!k_i x$. These spatial dependences are confirmed in \fig{warm}(d), even though, in computing the different curves, we have directly employed \eq{en2}, without any of the assumptions leading to Eqs.~\eqn{dgcurva}, \eqn{dgpartial} and \eqn{dggradb}. In particular, we notice that the heating associated with $\Delta \gamma_{\partial B}$ (blue line in \fig{warm}(d))  follows the profile of  magnetic energy shown in \fig{warm}(c), whereas the contribution of $\Delta \gamma_{\rm curv}$ (green line in \fig{warm}(d)) peaks --- though, with negative values --- at the same locations as the electric energy in  \fig{warm}(c). Lastly, the grad-B term $\Delta \gamma_{\grad B}$ (orange line in \fig{warm}(d)) is largest in between the peaks of $\delta B_\perp^2$ and $\delta E_\perp^2$ (e.g., see the orange line in \fig{warm}(d) at $3\lesssim x/r_{L,i}\lesssim 4.5$).


\section{Summary and Discussion}\label{sec:summary}
In this work, we have investigated the physics of electron heating during the growth of ion velocity-space instabilities in a collisionless plasma, by means of 1D and 2D fully-kinetic PIC simulations. We have focused on instabilities driven by a pressure anisotropy with $P_\perp>P_\parallel$: the mirror and ion cyclotron instability and the electron whistler instability.
We do not consider the initial value problem of the evolution of a prescribed temperature anisotropy. Rather, we constantly drive $P_\perp>P_\parallel$ by continuously forcing an amplification of the mean magnetic field. The increase in magnetic field, coupled to the adiabatic invariance of the particle magnetic moments, drives  $P_\perp> P_\parallel$. Our setup allows us to study the long-term non-linear evolution of the relevant instabilites, beyond their initial exponential stage.
In our setup, the increase in magnetic field is driven by compression.\footnote{As a side note, we remark that our numerical technique can also be applied to expanding plasmas, e.g., to study the interplay between electron and ion kinetic physics in the solar wind.} However, our results --- and in particular, our model of electron heating via the ion cyclotron instability ---  hold regardless of what drives the field amplification, so they can be equally applied to the case where velocity-space instabilities are induced by incompressible shear motions \citep{kunz_14,riquelme_14}.

Fully-kinetic PIC simulations with mobile ions and a realistic
mass ratio are required to capture from first principles the energy transfer from ions to electrons during the growth of ion velocity-space instabilities. To study this problem, we have modified the standard PIC equations (i.e., Maxwell's equations and the Lorentz force) to account self-consistently for the overall compression of our system. Anisotropy-driven {\it ion} instabilities in a system undergoing compression have been studied with a hybrid code by \citet{hellinger_05}. In hybrid codes, the electrons are treated as a massless charge-neutralizing isotropic fluid, so these codes cannot capture self-consistently the kinetic physics of anisotropy-driven {\it electron} instabilities (notably, the electron whistler mode), nor the efficiency of electron heating during the growth of ion-scale instabilities. 

With our novel fully-kinetic technique, we have studied, on the one hand, how the growth of ion-scale instabilities depends on the electron-to-ion temperature ratio $T_{0e}/T_{0i}$. On the other hand, we have assessed how the development of ion velocity-space instabilities can result in net electron heating (aside from the energy gain due to compression alone). Our main results are as follows:
\bi
\item For the values of ion plasma beta $\beta_{0i}\sim 5-30$ expected in the midplane of low-luminosity accretion flows \citep[e.g.,][]{sadowski_13}, the nature of the dominant anisotropy-driven ion instability changes as a function of the electron-to-ion temperature ratio. If $T_{0e}/T_{0i}\gtrsim 0.2$, the mirror instability dominates. This is in agreement with recent hybrid \citep{kunz_14} and PIC \citep{riquelme_14} simulations of shear-driven ion velocity-space instabilities in collisionless plasmas with $T_{0e}=T_{0i}$. For $T_{0e}/T_{0i}\gtrsim 0.2$, the electron anisotropy induced by compression actively participates in the growth of the mirror mode.\footnote{For warm electrons, the electron anisotropy exceeds the threshold of marginal stability for the electron whistler mode $[\beta_\perpe/\beta_\pare-1]_{\rm MS}\simeq \coeffe/\beta_\pare^\powe$ before the onset of the mirror instability. In this case, we find that electron pitch-angle scattering off the electron whistler waves maintains the electron anisotropy at the marginal stability threshold of the whistler mode.} In contrast, if the electron thermal content is reduced, the electron contribution to the excitation of the mirror instability is suppressed, and mirror modes are weaker. For $T_{0e}/T_{0i}\lesssim 0.2$, we find that it is the ion cyclotron mode that controls the evolution of the system, at least up to one compression timescale. Pitch-angle scattering off the resulting ion cyclotron waves reduces the ion anisotropy, which in the saturated  state follows the marginal stability condition $[\beta_\perpi/\beta_\pari-1]_{\rm MS}\simeq \coeffi/\beta_\pari^\powi$ appropriate for the ion cyclotron mode.
\item Since the wavevector of the ion cyclotron instability is aligned with the mean field, the relevant physics at temperatures $T_{0e}/T_{0i}\lesssim 0.2$ can be conveniently studied with 1D simulations, with the box aligned with the ordered field. In this regime, we find that the evolution of turbulent field energy and of the ion anisotropy in 1D simulations is in excellent agreement with 2D results.
Thanks to the greater number of particles per cell allowed by 1D simulations, the efficiency of electron heating by the ion cyclotron instability can thus be reliably estimated.
\item We have developed an analytical model to describe the physics of electron heating during the growth of the ion cyclotron instability, clarifying the physical origin of the different heating terms in Eqs.~\eqn{dgcurva}, \eqn{dgpartial}, \eqn{dggradb} and \eqn{dgexb}.\footnote{Our analytical model can be employed to describe the process of electron heating due to any ion-scale instability (i.e., not only the ion cyclotron mode) having transverse fluctuations as in  \eq{bfield} that are weak compared to the mean field, i.e., $\delta B_\perp/|\bavg|\ll1$.} We find that for cold electrons ($\beta_{0e}\lesssim 2 \,m_e/m_i$, where $\beta_{0e}$ is the ratio of electron thermal pressure to magnetic pressure), the electron energy gain is controlled by the magnitude of the E-cross-B velocity induced by the ion cyclotron waves (\eq{dgexb}). This term is independent of the initial electron temperature, so it provides a solid energy ``floor'' even for electrons starting with extremely low temperatures. On the other hand, the electron energy increase for $\beta_{0e}\gtrsim 2 \,m_e/m_i$  is proportional to the initial electron temperature, and depends linearly (in \eq{dgcurva} and \eq{dgpartial}) or quadratically (in \eq{dggradb}) on the energy  of the ion cyclotron fluctuations. We have validated our analytical model in both regimes, $\beta_{0e}\lesssim 2\, m_e/m_i$ and $\beta_{0e}\gtrsim 2\, m_e/m_i$, by means of 1D PIC simulations. We find excellent agreement with our analytical results for both the magnitude and the spatial dependence of the various heating terms.
In Paper II, we investigate how the electron heating efficiency by the ion cyclotron mode depends on flow conditions.
\ei
Our results have implications for the physics of  low-luminosity accretion flows, where the rate of Coulomb collisions in the innermost regions is extremely low, and the plasma is two-temperature. However, the same velocity-space instabilities that we discuss here play an important role in  the solar wind \citep{kasper_02,kasper_06,bale_09,maruca_11,maruca_12,matteini_07,matteini_13,cranmer_09,cranmer_12} and in the intracluster medium \citep{scheko_05,lyutikov_07,santos-lima_14}. So, our findings have implications for a variety of collisionless astrophysical plasmas.

A major outcome of our study is that, for the plasma parameters appropriate to the innermost regions of ADAFs, the mirror instability dominates for $T_{0e}/T_{0i}\gtrsim0.2$, whereas for $T_{0e}/T_{0i}\lesssim0.2$ the ion anisotropy is controlled by the ion cyclotron instability. In shearing box simulations of accretion flows that employ the so-called kinetic MHD
framework \citep{sharma06,sharma07}, the evolution of an anisotropic pressure tensor is incorporated in
the MHD equations of the plasma.  There, the physics of anisotropy-driven instabilities is included by setting a strict upper limit on the pressure anisotropy: once the pressure anisotropy exceeds the threshold value, pitch-angle scattering is expected to quickly reduce the anisotropy to its threshold value. Based on our findings, the threshold of marginal stability to be employed in kinetic MHD simulations should depend not only on the local ion fluid properties (notably, $\beta_{0i}$), but also on the electron-to-proton temperature ratio: the marginal stability for the mirror instability should be enforced when  $T_{0e}/T_{0i}\gtrsim0.2$, whereas the threshold condition for the ion cyclotron mode should be applied when  $T_{0e}/T_{0i}\lesssim0.2$. An assessment of the appropriate threshold condition to be used as a closure relation in fluid models is important for the large-scale dynamics of accretion flows, e.g., by affecting their angular momentum transport.

A second application of our work concerns the physics of electron heating in collisionless plasmas. The electron heating mechanism described in this work can be incorporated into global general-relativistic magnetohydrodynamic (GRMHD) simulations of low-luminosity accretion flows, by adding a source term to the electron thermal evolution. This will enable us to follow the electron thermodynamics as the electron fluid is heated by ion velocity-space instabilities or  cooled by radiation. It would help to clarify the equilibrium value of the electron-to-proton temperature ratio as a function of radius in collisionless accretion flows, and could be employed for modeling the radiative properties of systems like Sgr A$^*$ \citep[e.g.,][]{yuan_narayan_14}. A detailed assessment of the dependence of the electron heating efficiency on flow conditions is deferred to Paper II.

We conclude by discussing a few questions that our work has not addressed. We have not tackled the role of oblique compressions, i.e., when the compression is not orthogonal to the large-scale magnetic field. Such a study should be performed with multi-dimensional PIC simulations, to be presented elsewhere.

Finally, we remark that our investigation has been performed for homogeneous electron-proton plasmas, without heavier nuclei. It is known that the inclusion of a significant fraction of Helium nuclei may significantly raise the proton cyclotron threshold, thus suppressing the proton cyclotron instability, in favor of the mirror mode \citep{price_86}.  Also, spatio-temporal variations in the ambient magnetic field on  intermediate scales (larger than the proton Larmor radius, but smaller than the large-scale gradients in the disk) may affect the proton cyclotron instability more than the mirror instability. This is because the proton cyclotron instability relies on the cyclotron resonance condition, while the mirror instability is non-resonant. Intermediate-scale variations in the ambient field could shift the proton cyclotron mode out of resonance. An assessment of these issues is left for a future study.

\acknowledgements
 We thank J. Drake, L. Matteini, S. Naoz and especially R. Penna for useful discussions. L.S. is supported by NASA through Einstein
Postdoctoral Fellowship grant number PF1-120090 awarded by the Chandra
X-ray Center, which is operated by the Smithsonian Astrophysical
Observatory for NASA under contract NAS8-03060. This work is supported in part by NASA via the TCAN award grant NNX14AB47G. L.S. gratefully acknowledges access to the PICSciE-OIT High Performance Computing Center and Visualization
Laboratory at Princeton University. The simulations were also performed on XSEDE resources under
contract No. TG-AST120010, and on NASA High-End Computing (HEC) resources through the NASA Advanced Supercomputing (NAS) Division at Ames Research Center. 

\appendix
\section{A. The PIC Method in a Compressing Box}\label{sec:method}
Here, we describe the basic equations of the PIC method in a compressing or expanding box. We will  solve the relevant equations in the fluid {\it comoving} frame, which is related to the {\it laboratory} frame by a Lorentz boost, with velocity $\bmath{U}$. We only consider the non-relativistic limit $|\bmath{U}|/c\ll1$, as we justify below. Quantities measured in the laboratory frame will be labeled with the subscript ``L''. In the comoving frame, we adopt two sets of spatial coordinates, with the same time coordinate. The {\it unprimed} coordinate system in the fluid comoving frame has a basis of unit vectors, so it is the appropriate coordinate set to measure all  physical quantities. Below, it will be convenient to re-define the unit length of the spatial axes in the comoving frame such that a particle subject only to compression or expansion stays at fixed coordinates. This will be our {\it primed} coordinate system in the fluid comoving frame.

The location of a particle in the laboratory frame is related to its position in the primed coordinate system of the fluid comoving frame by $\bmath{x}_{\rm L}= \bmath{L}\, \bmath{x}'$, where compression or expansion are described by the diagonal matrix 
\be
\L=\frac{\partial \bmath{x}}{\partial \bmath{x}'}=
\left(
\begin{array}{ccc}
 a_x \,& 0 \,& 0 \\
 0\,& a_y \,& 0 \\
 0 \,& 0 \,& a_z \\
\end{array}\right)~~~,~~~\ell=\rm{det}(\bmath{L})~~~,
\ee
where $\ell$ is the determinant. In the case of compression perpendicular to an ordered magnetic field aligned with the $x$ direction, as assumed in the main body of the paper, we take 
\be\label{eq:aaa}
a_{x}=1~~~,~~~a_{y}=a_{z}=(1+q\,t')^{-1}~~.
\ee
In general, $a_x$, $a_y$ and $a_z$ are functions of time, but not of the spatial coordinates. Note that both primed and unprimed coordinate systems pertain to the same reference frame (the comoving frame), so they share the same time coordinate $t'=t$ (and so, $\ud t'=\ud t$).
 By differentiating $\bmath{x}_{\rm L}= \bmath{L} \, \bmath{x}'$, we find
\be\label{eq:transf1}
\ud\bmath{x}_{\rm L}&=&\bmath{L}\,\ud\bmath{x}'+\dot{\bmath{L}}\,\bmath{x}'\,\ud t'~~~,
\ee
 where $\dot{\bmath{L}}=\ud {\bmath{L}}/\ud t'$. From our discussion above, it follows that the relation between the primed and unprimed coordinate systems, both pertaining to the fluid comoving  frame, is such that $\ud \bmath{x}=\bmath{L}\,\ud\bmath{x}'$. By defining $\bmath{U}=\dot{\bmath{L}}\ \bmath{x}'=\dot{\bmath{L}}\bmath{L}^{-1}\,\bmath{x}_{\rm L}$, we can rewrite \eq{transf1} as 
 \be\label{eq:transf2}
\ud\bmath{x}_{\rm L}&=& \ud\bmath{x}+\bmath{U}\ud t~~~,
\ee
which describes the Lorentz transformation from the comoving frame to the laboratory frame, to first order in the boost velocity $|\bmath{U}|/c\ll1$. In a cosmological context, the velocity $\bmath{U}$ would correspond to the Hubble flow. In our setup, $U_x=0$, whereas $U_y=-q\,y'/(1+q\,t'^2)$, and similarly for $U_z$. Since the typical scale of our simulations is the ion Larmor radius $r_{L,i}$, we find $|U_y|/c\sim|U_z|/c\sim q \rli/c\sim(q/\omega_{0ci})(v_{{\rm th},i}/c)$, where $v_{{\rm th},i}$ is the ion thermal velocity. In the regime relevant for astrophysical accretion flows, we expect  non-relativistic ions ($v_{{\rm th},i}/c\ll1$) and slow compressions (i.e., $q/\omega_{0ci}\ll1$), so our assumption $|\bmath{U}|/c\ll1$ is easily satisfied.

In analogy to Eqs.~\eqn{transf1} and \eqn{transf2}, the Lorentz transformation of the time coordinate between the laboratory frame and the comoving frame is
\be
\!\!\!\!\!\!\!\!\!\!\ud t_{\rm L}&=&\ud t + (\bmath{U}/c^2)\cdot \ud \bmath{x}\\&=&\ud t' + (\bmath{U}/c^2)\cdot (\bmath{L}\,\ud\bmath{x}')~~,\label{eq:transf3}
\ee
to first order in the boost velocity $|\bmath{U}|/c\ll1$. 

From Eqs.~\eqn{transf1}-\eqn{transf3}, one can obtain the relation between the momentum of a particle in the laboratory frame and in the comoving frame, as well as the transformation of the particle Lorentz factor. We define $\bmath{p}_{\rm L}=m \,\ud \bmath{x}\lab/\ud \tau$ and $\bmath{p}'=m \,\ud \bmath{x}'/\ud \tau=\bmath{L}^{-1}\bmath{p}$, where $\tau $ is the proper time and $\bmath{p}=m \,\ud \bmath{x}/\ud \tau$ is the physical momentum in the unprimed coordinate system. From Eqs.~\eqn{transf1}-\eqn{transf3}, it follows that 
\be\label{eq:mom1}
\bmath{p}_{\rm L}&=&\bmath{L}\,\bmath{p}'+\bmath{U}\gamma'm\\&=&\bmath{p}+\bmath{U}\gamma\,m~~,\nonumber
\ee
and similarly, defining the particle Lorentz factor in the two frames as $\gamma\lab=\ud t\lab/\ud \tau$ and $\gamma'=\ud t'/\ud \tau=\gamma$, we find
\be\label{eq:mom2}
\gamma\lab &=&\gamma'+(\bmath{U}/c^2)\!\cdot \!(\L\,\bmath{p}'/m)\\&=&\gamma+(\bmath{U}/c^2)\!\cdot\! (\bmath{p}/m)~~,\nonumber
\ee
consistent with a Lorentz transformation at first order in $|\bmath{U}|/c\ll1$. It is trivial to show that $\gamma\lab=\sqrt{1+(p\lab/mc)^2}$ and $\gamma=\gamma'=\sqrt{1+(p/mc)^2}$ are consistent with \eq{mom2}  to first order in $|\bmath{U}|/c\ll1$.\footnote{We remark that it is the unprimed momentum $\bmath{p}$, and not the primed momentum $\bmath{p}'$, that appears in the expression for the Lorentz factor $\gamma=\gamma'$. As mentioned above, all physical quantities must be measured in the unprimed coordinate system.} The  equations above will be used to obtain the Lorentz force in the comoving frame.

From Eqs.~\eqn{transf1}-\eqn{transf3}, one can also derive the relation between the temporal and spatial derivatives, when transforming from the laboratory frame to the comoving frame (with primed coordinates). We obtain
\be\label{eq:der1}
\frac{\partial}{\partial \bmath{x}\lab}&=&\L^{-1}\frac{\partial}{\partial \bmath{x}'}-\frac{\bmath{U}}{c^2}\frac{\partial}{\partial t'}~~,\\
\frac{\partial}{\partial t\lab}&=&\frac{\partial}{\partial t'}-\bmath{U}\cdot \left( \L^{-1}\frac{\partial}{\partial \bmath{x}'}\right)~~,\label{eq:der2}
\ee
to first order in $|\bmath{U}|/c\ll1$. The two equations above will be used to find the form of Maxwell's equations in the comoving frame, as we describe below.

Before proceding, we list the assumptions entering our derivation below of Maxwell's equations and the Lorentz force in the fluid comoving frame:
\bi
\item We assume that the compression or expansion velocity is non-relativistic, i.e., $|\bmath{U}|/c\ll1$. So, we retain only the terms that depend linearly on $|\bmath{U}|/c$, neglecting higher order terms, e.g., proportional to $\bmath{U}^2/c^2\propto \dot{\L}^2$. For the same reason, we neglect terms containing $\ddot{\L}\dot{\L}$, since this is the temporal derivative of $\dot{\L}^2/2$, which we consistently discard.
\item We allow for the possibility that the rate of expansion or compression of the system may change with time, i.e., $\ddot{\L}\ne 0$. Acceleration terms proportional to $\ddot{\L}$ are retained in the equations below, but we argue that they can be neglected in all the circumstances relevant for this work. In addition, we remark that our conclusions are the same when we adopt a different temporal evolution for the matrix $\L$, such that $\ddot{\L}=0$ (e.g., if we choose $a_y=a_z=1-q\,t$, as opposed to our standard choice $a_y=a_z=\qt^{-1}$). So, our main results hold regardless of whether the compression (or  expansion) speed is constant or changing over time. 
\item The Lorentz force that we obtain below in the fluid comoving frame holds for any particle momentum, i.e., for non-relativistic, trans-relativistic or ultra-relativistic particles.
\ei

\subsection{Maxwell's Equations}
From the usual expressions of Maxwell's equations in the laboratory frame, and using Eqs.~\eqn{der1} and \eqn{der2}, we find the form of Maxwell's equations in the comoving frame, to first order in $|\bmath{U}|/c\ll1$:
\be
\!\!\!\!\!\!\!\!\!\!\!\!\!\!\!\!\!\!\!\nabla'\!\cdot\!(\ell\, \bmath{L}^{-1} \bmath{E})&=&4 \pi\, \ell\, \rho' ~,\label{eq:eq1}\\
\!\!\!\!\!\!\!\!\!\!\!\!\!\!\!\!\!\!\!\nabla'\!\cdot\!(\ell\, \bmath{L}^{-1} \bmath{B})&=&0~,\label{eq:eq2}\\
\!\!\!\!\!\!\!\!\!\!\!\!\!\!\!\!\!\!\!\nabla'\!\cross\!(\L \bmath{E})  &=&-\frac{1}{c}\frac{\partial}{\partial t'}(\ell\, \bmath{L}^{-1} \bmath{B})\!-\!\frac{\ddot{\L}\bmath{x}'}{c^2}\!\cross\! \evec~,\label{eq:eq3}\\
\!\!\!\!\!\!\!\!\!\!\!\!\!\!\!\!\!\!\!\nabla'\!\cross\!(\L \bmath{B}) &=&\frac{1}{c}\frac{\partial}{\partial t'}(\ell\, \bmath{L}^{-1} \bmath{E})+\frac{4\pi}{c} \ell\, \bmath{J}'\!-\!\frac{\ddot{\L}\bmath{x}'}{c^2}\!\cross\! \bvec~,\label{eq:eq4}
\ee
where the temporal and spatial derivatives pertain to the primed coordinate system (the primed and unprimed systems share the same time coordinate, so $\partial/\partial t'=\partial/\partial t$, whereas spatial derivatives differ: $\nabla'=\L \,\nabla$). We define $\evec$ and $\bvec$ to be the physical electromagnetic fields measured in the unprimed coordinate system, which are related to the fields in the laboratory frame via
\be
\evec=\evec\lab +\frac{\bmath{U}}{c}\cross \bvec\lab\simeq \evec\lab +\frac{\bmath{U}}{c}\cross \bvec~,\\
\bvec=\bvec\lab -\frac{\bmath{U}}{c}\cross \evec\lab\simeq \bvec\lab -\frac{\bmath{U}}{c}\cross \evec~,
\ee
consistent with a Lorentz transformation at first order in $|\bmath{U}|/c\ll1$.  At the same order, the charge and current density transform as
\be
\rho\lab&=&\rho'+(\bmath{U}/c^2)\cdot (\L \,\bmath{J}')~,\\
\bmath{J}\lab&=&\L \,\bmath{J}'+\bmath{U}\rho'~,
\ee
and the primed quantities $\rho'$ and $\bmath{J}'$ (measured in the comoving frame with respect to the primed coordinate system) appear in Eqs.~\eqn{eq1} and \eqn{eq4}, respectively. In the unprimed coordinate system, $\rho=\rho'$ and $\bmath{J}=\L\,\bmath{J}'$. The form of the electric charge $\rho'$ and of the electric current density $\bmath{J}'$  will be discussed below.

In the following, we will neglect the two acceleration terms ($\propto \ddot{\L}$) present in Eqs.~\eqn{eq3} and \eqn{eq4} (we have already neglected terms containing $\ddot{\L}\dot{\L}$). Since the typical scale of our simulations is the ion Larmor radius, the characteristic magnitude of the  acceleration term in \eq{eq4} will be $\sim q^2 \rli B_0/c^2$, where $B_0$ is the strength of the ordered magnetic field. This should be compared with the left hand side of \eq{eq4}, which is of order $\sim \delta B_{\perp}/\rli$, where $ \delta B_{\perp}$  is the characteristic magnitude of the trasverse fluctuations associated with the ion cyclotron instability. In the main body of the text, we find that $\delta B_\perp/B_0\sim1$ after the exponential growth of the ion cyclotron mode, so that the acceleration term in \eq{eq4} can be neglected if $(q/\omega_{0ci})^2(v_{{\rm th},i}/c)^2\ll1$, which is certainly satisfied in our study.\footnote{As a side note, we point out that for our 1D simulations where the box is along the uncompressed direction $x$, the two acceleration terms in Eqs.~\eqn{eq3} and \eqn{eq4} vanish, since $\ddot{a}_x=0$.}
A similar argument shows that the acceleration term in \eq{eq3} can be neglected, since $(q/\omega_{0ci})^2(v_{{\rm th},i}/c)^2\ll1$. Moreover, as anticipated above, we have explicitly checked that our conclusions still hold when we adopt a different temporal evolution for $\L$ with $\ddot{\L}=0$ (viz., $a_x=1$, $a_y=a_z=1-q\,t$, as opposed to our choice in \eq{aaa}).

Our electromagnetic PIC algorithm solves for the two evolutionary equations \eqn{eq3} and \eqn{eq4}, neglecting the acceleration terms. The charge-conserving algorithm implemented in TRISTAN-MP, which ensures that
\be\label{eq:chg3}
\frac{\partial}{\partial t'}(\ell\,\rho'\,c)+\nabla'\cdot(\ell\, \bmath{J}')  &=&0~~~,
\ee
will automatically satisfy \eq{eq1} at all times, provided that it is satisfied at the initial time. 

The time-dependent terms in the matrix $\bmath{L}$ and in its determinant $\ell$ affect the Courant--Friedrichs--Lewy condition for numerical stability. For the sake of simplicity, let us consider the case of isotropic compression $a_x=a_y=a_y\equiv a$ and let us assume that the characteristic compression rate is much smaller than the frequency of electromagnetic waves. The Courant--Friedrichs--Lewy condition dictates that
\be\label{eq:cfl}
c\le\frac{a}{\sqrt{3}}\frac{\Delta x'}{\Delta t'}~~~,
\ee
for a three-dimensional computational box with $\Delta x'=\Delta y'=\Delta z'$ (for a two-dimensional domain, one should replace $\sqrt{3}$ with $\sqrt{2}$). For $a=\qt^{-1}$, any given choice of the numerical speed of light will eventually run into stability issues. Yet, since we are interested in the evolution of the system only for $\qt\lesssim 2$, we typically choose $c=0.15$ cells/timestep, which is small enough to avoid stability issues. We have checked that a different choice of the numerical speed of light ($c=0.2$ rather than $c=0.15$) yields identical results.\footnote{For an expanding system, $a\ge1$ and so any choice of $c\le\Delta x'/\sqrt{3}\Delta t'$ will ensure that the system is stable at all times.}

The numerical implementation of \eq{eq3} and \eq{eq4} is straightforward, once we have confirmed that we may neglect the acceleration terms. In TRISTAN-MP, the electric and magnetic fields are leap-frogged in time, so that the algorithm is second order accurate. Here, it is convenient to define
\be
\label{eq:bcal}
\bmath{B}'\equiv \ell\, \bmath{L}^{-1} \bmath{B}~~~~~~~~~ \bmath{E}'\equiv \ell\, \bmath{L}^{-1} \bmath{E}~~~,
\ee
so that the two evolutionary  equations become 
\be
\nabla'\cross(\ell^{-1}\L^2 \bmath{E}')  &=&-\frac{1}{c}\frac{\partial \bmath{B}'}{\partial t'} \label{eq:eq3b}~~,\\
\nabla'\cross(\ell^{-1}\L^2 \bmath{B}') &=&\frac{1}{c}\frac{\partial \bmath{E}'}{\partial t'} +\frac{4\pi}{c} \ell\, \bmath{J}'~~,\label{eq:eq4b}
\ee
and the algorithm is still second order accurate in time, provided that the term $\ell^{-1}\L^2$ is evaluated at the same time as $\bmath{E}'$ in \eq{eq3b} and at the same time as $\bmath{B}'$ in \eq{eq4b}. From \eq{eq3b}, it is easy to derive the expected evolution of an ordered magnetic field, as a result of compression. In the case explored in the main text, $a_x=1$, $a_y=a_z=\qt^{-1}$, an ordered field aligned with the $x$ direction will evolve such that $\bmath{B}'=\ell\, a_x^{-1} \bmath{B}=$const, or $ \bmath{B}\propto \qt^2$. This is indeed the scaling expected from flux freezing. Similarly, if the ordered field were to be aligned with $y$ or $z$, it would evolve as $ \bmath{B}\propto \qt$, which is again compatible with flux freezing.

\subsection{The Lorentz Force}
The Lorentz force in the comoving frame can be obtained from the Lorentz force in the laboratory frame by differentiating \eq{mom1} with respect to time, and remembering that $\ud t\lab/\ud t'=\gamma\lab/\gamma'$, which can be obtained from \eq{mom2}. At first order in $|\bmath{U}|/c\ll1$, we find
\be
\label{eq:eql0}
\!\!\!\!\!\frac{\ud \bmath{p}'}{\ud t'}&=&-2 \,\dot{\L} \L^{-1} \bmath{p}'+q\, \L^{-1}\left(\evec+\frac{\L \bmath{v}'}{c}\cross \bvec\right)\\&&-\ddot{\L}\L^{-1}\bmath{x}'\gamma'~~,\nonumber
\ee
where $q$ is the particle charge and we have defined $\bmath{v}'=\ud \bmath{x}'/\ud t'=\bmath{p'}/\gamma'm$. The first term on the right hand side in \eq{eql0} accounts for the overall compression or expansion of the system. In the following, we will neglect the acceleration term $\propto \ddot{\L}$ (we have already neglected terms containing $\ddot{\L}\dot{\L}$). By comparison with the first term on the right hand side of \eq{eql0}, the acceleration term can be discarded if $|\bmath{v}'|\gg q\,\rli$. For ions, taking $|\bmath{v}'|\sim v_{{\rm th},i}$ this is equivalent to $q/\omega_{0ci}\ll1$, which is fulfilled. For electrons, it constrains the electron-to-ion temperature ratio to be $T_{0e}/T_{0i}\gg (q/\omega_{0ci})^2$, which is realized in all the cases explored in this work.

By neglecting the acceleration term,  the evolution of the particle orbits in the primed coordinate system is controlled by the following set of coupled equations,
\be\label{eq:eql1}
\!\!\!\!\!\!\frac{\ud \bmath{p}'}{\ud t'}&=&-2 \,\dot{\L} \L^{-1} \bmath{p}'+q\, \L^{-1}\left(\evec+\frac{\L \bmath{v}'}{c}\cross \bvec\right)~,\\
\!\!\!\!\!\!\frac{\ud \bmath{x'}}{\ud t'}&=&\bmath{v}'~.\label{eq:eql2}
\ee
 In our implementation of the Lorentz force, we do not solve \eq{eql1}, but rather the corresponding equation for the momentum $\bmath{p}$ in the unprimed coordinate system,
\be
\frac{\ud \bmath{p}}{\ud t'}&=&-\,\dot{\L} \L^{-1} \bmath{p}+q\left(\evec+\frac{\bmath{v}}{c}\cross \bvec\right)\label{eq:eql1b}~~~,
\ee 
which is simpler to implement (see below), and has the advantage that $\bmath{p}$ and $\bmath{v}=\bmath{p}/\gamma\, m=\L^{-1}\bmath{v}'$ correspond to the physical momentum and velocity in the fluid comoving frame, respectively.

 The equation for the evolution of the particle energy follows from \eq{eql1b}. With the same approximations adopted above, we find
\be
\frac{\ud \gamma m c^2}{\ud t'}=-(\dot{\L} \,\bmath{p})\cdot(\L^{-1}\bmath{v})+q \,\bmath{v}\cdot \evec~~~.
\ee
In summary, we solve \eq{eql2} and \eq{eql1b}, with the primed velocity in \eq{eql2} being $\bmath{v}'=\L^{-1}\bmath{v}$. We remark that the fields $\evec$ and $\bvec$ in  \eq{eql1b} are the electromagnetic fields in the unprimed coordinate system, related to $\bmath{B}'$ and $\bmath{E}'$ via \eq{bcal}.

In the case explored in the main text, viz., $a_x=1$, $a_y=a_z=\qt^{-1}$ (see \eq{aaa}), the Lorentz force in \eq{eql1b} gives the correct result that, in the absence of electromagnetic fields, the particle momentum in a direction that does not suffer compression is unchanged, whereas the particle momentum in a direction that gets compressed increases as $p_\perp/p_{0\perp}=(1+q\,t)$. In the presence of an ordered magnetic field $\bvec$ aligned with the non-compressed axis (as in \fig{setup}, where the field is along $x$), the equation for the perpendicular momentum $p_{\perp}=(p_y^2+p_z^2)^{1/2}$ can be obtained by summing the equations for $p_y$ and $p_z$, after mutiplying them by $p_y$ and $p_z$, respectively. We obtain 
\be
\frac{1}{2}\frac{\ud p_{\perp}^2}{\ud t}=\frac{q}{1+q\,t}\,p_{\perp}^2~~~,
\ee
which yields again $p_\perp/p_{0\perp}=\qt$, as in the case without any magnetic field. On the other hand, the parallel momentum does not change during compression, so $p_{\parallel}=p_{0\parallel}$. The conservation of the first ($\mu\propto p_{\perp}^2/| \bvec|$) and second ($J\propto p_{\parallel} |\bvec|/n$) adiabatic invariants easily follows, since $n\propto \qt^2$ and $|\bvec|\propto \qt^2$. So, our equations are consistent with the so-called CGL double adiabatic model of \citet{chew_56}. Also, the temporal evolution of $p_{\perp}$ and $p_\parallel$ justifies our choice of quantifying the electron heating in terms of the $\chi$ parameter introduced in \sect{struct1}.
 
We conclude by describing our numerical implementation of \eq{eql2} and \eq{eql1b}, i.e., how to solve for the unprimed momentum $\bmath{p}$ and the primed position $\bmath{x}'$. In TRISTAN-MP, the particle momenta and coordinates are leap-frogged in time. To preserve second-order accuracy, the time-dependent matrix $\L$, which is needed for $\bmath{v}'=\L^{-1}\bmath{v}$ in \eq{eql2}, should be evaluated at the same time as the particle velocity in \eq{eql2}. 

As regard to \eq{eql1b}, in the standard Boris pusher employed in most PIC codes, the update of the particle momentum is split into three steps: (\textit{i}) acceleration via the electric field for half a timestep; (\textit{ii}) rotation by the magnetic field for a full timestep; (\textit{iii}) acceleration via the electric field for half a timestep. The rotation by the magnetic field is still $\bmath{v}\cross\bvec$, i.e., it is unchanged by the compression or expansion of the box. We now describe how steps (\textit{i}) and (\textit{iii}) need to be modified, to preserve second order accuracy. Let $\bmath{p}_n$ and $\bmath{p}_{n+1}$ be the particle momenta at timestep $n$ and $n+1$, respectively,  and let $\bmath{p}_{ n^-}$ and $\bmath{p}_{ n^+}$ be the momenta respectively before and after the magnetic rotation. We account for both the electric acceleration and the overall compression (or expansion) of the box by solving
\be
\!\!\!\!\!\!\!\!\!\!\!\!\!\!\!\!\frac{p_{i,n^-}\!-p_{i,n}}{\Delta t/2}\!&=&-\!\left(\frac{\dot{a}_i}{a_i}\right)_{\!n+\frac{1}{2}}\!\!\!\!\!\!\frac{p_{i,n}\!+\!p_{i,n^-}}{2}\!+\!q E_{i,n+\frac{1}{2}}~,\\
\!\!\!\!\!\!\!\!\!\!\!\!\!\!\!\!\frac{p_{i,n+1}\!-p_{i,n^+}}{\Delta t/2}\!&=&-\!\left(\frac{\dot{a}_i}{a_i}\right)_{\!n+\frac{1}{2}}\!\!\!\!\!\!\frac{p_{i,n^+}\!+\!p_{i,n+1}}{2}\!+\!q E_{i,n+\frac{1}{2}}~,
\ee
where $i=x,\,y$ or $z$. The first equation is used to derive $p_{i,n^-}$, and from the latter we obtain $p_{i,n+1}$. An alternative implementation relies on the fact that the vector $\L\, \bmath{p}$ is constant during the compression (or expansion) of the box, as one can easily infer from \eq{eql1b} in the case $\evec=\bvec=0$. We have checked that our results are the same for these two different implementations of the Lorentz force.
 
\subsection{The Electric Current Density}
The self-consistent interplay between particles and electromagnetic fields at the basis of the PIC method will be completed once we prescribe the electric current density $\bmath{J}'$ as a sum of the contributions of individual particles. In the unprimed coordinate system of the fluid comoving frame, the electric current density is simply
\be
\bmath{J}(\bmath{x},t)=\sum_{\alpha} q_\alpha \bmath{v}_\alpha S[\bmath{x}-\bmath{x}_\alpha(t)]~~~,
\ee
where the summation is over the particles $\alpha$, whose charges, positions and velocities are respectively $q_\alpha$, $\bmath{x}_\alpha$ and  $\bmath{v}_\alpha$. The weighting function $S$ depends on the choice of the charge deposition scheme. The electric current density $\bmath{J}'$ in the primed system is
\be\label{eq:chg1}
\!\!\!\!\!\!\!\!\!\bmath{J}'=\L^{-1}\bmath{J}=\ell^{-1}\!\!\sum_{\alpha} q_\alpha \bmath{v}'_\alpha S[\bmath{x}'-\bmath{x}'_\alpha(t')]~~~,
\ee
where we have used the fact that $\bmath{v}'=\L^{-1}\bmath{v}$ and that the determinant of the coordinate transformation is $|\partial \bmath{x}'/\partial \bmath{x}|=\ell^{-1}$. It follows that the source term in \eq{eq4} reads
\be
\!\!\!\frac{4\pi}{c}\ell\,\bmath{J}'=\frac{4\pi}{c}\sum_{\alpha} q_\alpha \bmath{v}'_\alpha S[\bmath{x}'-\bmath{x}'_\alpha(t')]~~~.
\ee
Similarly, the charge density in the primed system is
\be\label{eq:chg2}
\!\!\!\!\!\!\!\!\!\rho'=\ell^{-1}\!\!\sum_{\alpha} q_\alpha S[\bmath{x}'-\bmath{x}'_\alpha(t')]~~~.
\ee
From \eq{chg1} and \eq{chg2}, the charge conservation in \eq{chg3} readily follows.

\section{B. The Electron Whistler Instability}\label{sec:lecs}
Here, we summarize the main properties of the electron whistler instability, which has been widely discussed in the literature \citep[e.g.,][]{gary_96,gary_99,gary_00b,nishimura_02,gary_11}. If the electron anisotropy exceeds the threshold in \eq{marge}, the whistler instability generates right-handed transverse electromagnetic waves propagating along or opposite to the mean magnetic field. The dominant wavelength at marginal stability is $\lambda_e= 2\pi A_{e,\rm MS}^{-1/2}\,c/\omega_{\rm pe}$ and the oscillation frequency is $\omega_e= A_{e,\rm MS}\,(A_{e,\rm MS}+1)^{-1} \omega_{ce}$ \citep{kennel_66,yoon_87,yoon_11,bashir_13}. Since the electron anisotropy at marginal stability is $A_{e,\rm MS}=[\beta_\perpe/\beta_\pare-1]_{\rm MS}\propto \beta_{\pare}^{-\powe}$ (see \eq{marge}), the dominant wavelength scales as $\lambda_e\propto \beta_\pare^{0.25}$, as verified by \citet{gary_06} by means of PIC simulations.

The whistler mode is not the only instability that grows on electron scales if $\beta_\perpe>\beta_\pare$. However, it is the dominant  mode in the parameter range that we have explored in this work. In a compressing system, electron instabilities will be able to grow in a pristine homogeneous plasma only if they are excited before the onset of ion-driven instabilities (either the mirror or the ion cyclotron). Since the ion anisotropy reaches at most $A_i\sim2$ before ion modes go unstable (e.g., see \fig{1d2d}(a) or \fig{fluid1d}(a)), a necessary condition for electron instabilities to develop in our driven setup is that the electron anisotropy at marginal stability in \eq{marge} is
\be\label{eq:whis}
A_{e,\rm MS}=\frac{S_e}{\beta_{\pare}^{\alpha_e}}\lesssim 2~~~,
\ee
which implies $\beta_{\pare}\gtrsim0.1$ for $S_e\simeq\coeffe$ and $\alpha_e\simeq\powe$. We now show that, in this range of electron beta, the whistler instability dominates.

As discussed by \citet{gary_06}, the electron mirror instability, akin to the ion mirror instability presented in \sect{struct2}, is a purely growing mode with compressive fluctuations, having maximum growth rate at propagation oblique to the mean field.  \citet{gary_06} have shown that for $0.1\lesssim \beta_{e\parallel}\lesssim 1000$, the whistler instability has a larger growth rate and a lower threshold than the electron mirror. So, the electron mirror will never dominate in the parameter regime explored in our work.

Competing instabilities will be excited in low-beta plasmas with strong magnetizations, i.e., $\omega_{ce}\gtrsim \omega_{\rm pe}$. For $\beta_\pare\lesssim 0.025$ and $ \omega_{ce}\sim \omega_{\rm pe}$, the maximum growth rate is at propagation oblique to the mean field \citep{gary_99,gary_00b,gary_11}, and the fluctuating electric fields become predominantly electrostatic (as opposed to the electron mirror mode, which is also oblique). This electrostatic mode is never important in our parameter range, since $\beta_\pare\gtrsim 0.1$ from \eq{whis}. In fact, the waves that are excited by the electron anisotropy in our simulations, both in 2D (\fig{fluid2d}) and in 1D (\fig{fluid1d}), do not display any significant electrostatic component.
 
Finally, the Z-mode instability should become important in highly magnetized plasmas, if $\omega_{ce}\gtrsim 2\,\omega_{\rm pe}$ \citep{gary_99,gary_00b,gary_11}. Including relativistic effects, the proper definitions of the electron cyclotron frequency and plasma frequency are $\omega_{ce}=e|\bavg|/\la\gamma_e\ra m_e c$ and $\omega_{\rm pe}=\sqrt{4 \pi n e^2/\la \gamma_e \ra m_e}$, respectively smaller by a factor of $\la \gamma_e\ra$ and of $\la \gamma_e\ra^{1/2}$ than the non-relativistic formulae. Here $\la \gamma_e\ra$ is the average electron Lorentz factor. The condition $\omega_{ce}\gtrsim 2\,\omega_{\rm pe}$ for the growth of the Z-mode is equivalent to
\be\label{eq:fin1}
\frac{v_{Ai}}{c}\sqrt{\frac{m_i}{\la \gamma_e\ra m_e}}\gtrsim 2~~~.
\ee
In the ultra-relativistic limit $\la \gamma_e\ra\simeq 3\,k_B T_e/m_e c^2$, so the condition in \eq{fin1} for the dominance of the Z-mode instability is equivalent to $\beta_e\lesssim0.1$, which is never satisfied in our parameter range, due to \eq{whis}. For non-relativistic electrons, \eq{fin1} can be rewritten as 
\be
\frac{k_B T_{e}}{m_e c^2}\frac{1}{\beta_e}\gtrsim 2~~~.
\ee
The assumption of non-relativistic electrons then is equivalent to $\beta_e\ll 0.5$, which is generally not satisfied in our setup, if we require that electron-scale instabilities 
precede ion-scale instabilities (see \eq{whis}).

\bibliography{bimax}
\end{document}